# Hydrated Peridotite – Basaltic Melt Interaction Part I: Planetary Felsic Crust Formation at Shallow Depth

Anastassia Y. BORISOVA[1,2*], Nail R. ZAGRTDENOV[1], Michael J. TOPLIS[3], Wendy A. BOHRSON[4], Anne NEDELEC[1], Oleg G. SAFONOV[2,5,6], Gleb S. POKROVSKI[1], Georges CEULENEER[1], Ilya N. BINDEMAN[7], Oleg E. MELNIK[8], Klaus Peter JOCHUM[9], Brigitte STOLL[9], Ulrike WEIS[9], Andrew Y. BYCHKOV[2], Andrey A. GURENKO[10], Svyatoslav SHCHEKA[11], Artem TEREHIN[5], Vladimir M. POLUKEEV[5], Dmitry A. VARLAMOV[5], Kouassi E.A. CHARITEIRO[1], Sophie GOUY[1], Philippe de PARSEVAL[1]

[1] *Géosciences Environnement Toulouse, Université de Toulouse; UPS, CNRS, IRD, Toulouse, France*

[2] *Geological Department, Lomonosov Moscow State University, Vorobievy Gory, 119899, Moscow, Russia*

[3] *Institut de Recherche en Astrophysique et Planétologie (IRAP) UPS, CNRS, Toulouse, France*

[4] *Central Washington University, Department of Geological Sciences, Ellensburg, WA 98926, USA*

[5] *Korzhinskii Institute of Experimental Mineralogy, 142432, Chernogolovka, Moscow region, Russia*

[6] *Department of Geology, University of Johannesburg PO Box 524, Auckland Park, 2006, Johannesburg, South Africa*

[7] *Geological Sciences, University of Oregon, 1275 E 13th street, Eugene, OR, USA*

[8] *Institute of Mechanics, Moscow State University, 1- Michurinskii prosp, 119192, Moscow, Russia*

[9] *Climate Geochemistry Department, Max Planck Institute for Chemistry, P.O. Box 3060, D-55020 Mainz, Germany*

[10] *Centre de Recherches Pétrographiques et Géochimiques, UMR 7358, Université de Lorraine, 54501 Vandœuvre-lès-Nancy, France*

[11] *Bavarian Research Institute of Experimental Geochemistry and Geophysics (BGI), University of Bayreuth, 95440 Bayreuth, Germany*

*Corresponding author: E-mail: **anastassia.borisova@get.omp.eu**;
*Corresponding address: Géosciences Environnement Toulouse UMR 5563, Observatoire Midi Pyrénées, 14 Avenue E. Belin, 31400 Toulouse, France; Tel: +33(0)5 61 54 26 31; Fax: +33(0)5 61 33 25 60

# ABSTRACT

Current theories suggest that on Earth and, possibly, on other terrestrial planets early in their history, the first continental crust may has been produced by direct melting of hydrated peridotite. However, the conditions, mechanisms and necessary ingredients of such production remain elusive. To fill this gap, we have conducted experiments of serpentinite melting in the presence of variable proportions (from 0 to 87 wt%) of basaltic melt, at typical conditions of the shallow lithosphere and asthenosphere: 1250 to 1300°C, and 0.2 to 1.0 GPa. These experiments revealed formation of silica-rich liquids (up to 66 – 71 wt% $SiO_2$) at 0.2 GPa, which are similar to tonalite-trondhjemite-granodiorite magmas (TTG) identified in modern terrestrial oceanic mantle settings. By contrast, at 1.0 GPa silica-poorer liquids (≤ 51 wt% $SiO_2$) were formed. Our results suggest a new mechanism of aqueous fluid-assisted partial melting of peridotite that may have operated on the early Earth and Mars just after the solidification of an ultramafic-mafic magma ocean, at depths of less than 10 km, leading to the formation of the first embryos of continental crust. The proposed mechanism of the continental crust formation may have been predominant before the onset of plate tectonics.

# INTRODUCTION

The conditions and mechanisms that led to the production of the earliest intermediate to felsic (Si- and Al-enriched) crust on Earth and Mars are the subject of intense debate (Rudnick and Gao, 2003; Harrison, 2009; Reimink et al., 2014; 2016; Sautter et al., 2015; 2016; Burnham and Berry, 2017). There is currently a consensus that such crust was tonalite-trondhjemite-granodiorite (TTG) in composition, not only in the Archean, but possibly as far back as the geological record goes (e.g. the parental melt of the 4.37 – 4.02 Ga Jack Hills detrital zircons, Burnham and Berry, 2017). Ancient granodioritic rocks have also been identified on Mars (Sautter et al., 2015; 2016), indicating that formation of felsic crust also occurred on other terrestrial planets even without plate tectonics and subduction. Dry peridotite melting at shallow levels does not produce appropriate melt compositions (Rudnick and Gao, 2003; Harrison, 2009; Reimink et al., 2014; 2016; Burnham and Berry, 2017), thus most currently accepted models for the generation of Earth's and Martian felsic crust consider a two- or multi-stage process, which involves extraction of basaltic magma from the peridotite mantle followed by its fractional crystallization (Reimink et al., 2014; 2016; Udry et al., 2018) and/or re-melting of hydrated mafic rocks at deep lithospheric conditions (Harrison, 2009; O'Neil and Carlson, 2017). However, the higher mantle temperature in the Hadean allows us to envision petrological mechanisms that are not typical for felsic magma generation today (i.e. that do not require subduction, Herzberg et al., 2010; Sautter et al., 2016). For example, the early continental crust of intermediate to felsic composition on Earth might have been created by direct melting of a hydrated peridotite at shallow depths (Rudnick, 1995; Rudnick and Gao, 2003). A possible present-day analogue is the formation of tonalites-trondhjemites in the shallow mantle beneath oceanic spreading centers (called oceanic plagiogranites in this context) (Coleman and Peterman, 1975; Amri et al., 1996; 2007). Indeed, low pressure melting experiments ($\leq 1.0$ GPa) involving peridotite and high water contents indicate that peridotite-derived partial melts may become enriched in $SiO_2$ and other lithophile elements (Al, alkalis), with abundances approaching those of the continental crustal rocks (Hirschmann et al., 1998; Ulmer, 2001). The increase in the silica contents of partial melts at low pressures is thought to be related to combined effects of water, alkalis and pressure on the structure of the aluminosilicate melt, shifting olivine-pyroxene cotectics to higher $SiO_2$ contents through an increase in the activity coefficient of silica (Hirschmann et al., 1998).

In this low-pressure context, the question arises to what extent basaltic melt has a direct (through reaction with its host peridotite) or indirect (heat source) impact on the production of felsic crust. In this respect, experiments that reproduce basaltic melt-peridotite systems at lithospheric to asthenospheric pressures (<0.8 GPa) are currently limited to nearly anhydrous conditions (Fisk, 1986; Kelemen et al., 1990; Morgan, and Liang, 2003; Van den Bleeken et al., 2010; 2011). While dry basaltic to andesitic-basaltic melts in association with olivine were observed in this case, no data exist for the case of partial melting of hydrated peridotite (or serpentinite) due the emplacement of basaltic melt at shallow crustal depths (7 – 30 km). This latter case is of primary interest for the early Earth (and probably Mars), as ultramafic-mafic protocrust formed from magma ocean crystallization was most probably hydrated, either from volatiles released by magma ocean crystallization or through interaction with surface liquid water (Albarede and Blichert-Toft, 2007; Elkins-Tanton, 2012). A present-day analogue is the level of the mantle and the petrologic MOHO transition in the oceanic lithosphere, where interaction of basaltic magmas with peridotites produces chromitite-dunite associations (Kelemen et al., 1995; Arai, 1997; Borisova et al., 2012; Zagrtdenov et al., 2018; Rospabé et al., 2019), and the hydrated peridotite partial melting (Benoit et al., 1999). Such shallow conditions of hydrated peridotite melting in the presence of basaltic melt might also occur in mantle plumes (Reimink et al., 2014; 2016; Bindeman, 2008), Hadean heat-pipe volcanoes (Moore and Webb, 2013), during meteoric impacts (Marchi et al., 2014) and during magma ocean solidification (Elkins-Tanton, 2012). For all these reasons, we have conducted a set of new experiments in a pure serpentinite and hybrid serpentinite-basalt systems at low pressure (≤1 GPa) to investigate the generation of intermediate to felsic crust. Modern (i.e. post-Archean) magmatism near the petrologic MOHO in the oceanic lithosphere will be used as a ground-truth test for the processes proposed at the light of our experiments.

# EXPERIMENTAL STRATEGY, METHODS AND TECHNIQUES

## Experimental Strategy

The experiments in the hybrid and mixed system containing serpentinite and variable proportion of basalt (**Supplementary Table A1**) were performed at 0.2 to 1.0 GPa and 1250 to 1300°C. Although in modern oceanic settings, temperatures of 1050°C are sufficient to initiate reaction of hydrated peridotite with basaltic magma at 0.2 GPa (Borisova et al., 2012), we chose higher temperatures for the experiments as they provide a closer analogue of the thermal conditions of the upper mantle as well as for the Hadean and early Martian mantle which are believed to be hotter than the modern mantle (Herzberg et al., 2010; Sautter et al., 2016). In addition, higher temperatures substantially increase reaction rates of the serpentinite-basaltic melt interactions and ensure conditions corresponding to complete melting of the basalt, consistent with the majority of existing models of basaltic melt extraction from the mantle (Fisk, 1986; Hirschmann et al., 1998; Ulmer, 2001; Morgan and Liang, 2003).

## Experimental Techniques

Three types of experiments were performed: (*i*) pure serpentinite dehydration, (*ii*) mixed runs on the serpentinite-basaltic melt interaction and (*iii*) hybrid experiment of the serpentinite-basaltic melt interaction (**Supplementary Table A2**). Two piston-cylinder apparatuses were used in our experiments. A part of hybrid experiments (P1, P3, P7) at 0.5 – 1.0 GPa and 1300°C used the end-loaded Boyd-England piston-cylinder apparatus at the Korzhinskii Institute of Experimental Mineralogy, Chernogolovka, Russia (Safonov et al., 2014). The second part of hybrid experiments (P15, P18) at 0.5 GPa and 1300°C used the "Max Voggenreiter" end-loaded Boyd-England piston-cylinder apparatus at the Bavarian Research Institute of Experimental Geochemistry and Geophysics (BGI), Bayreuth, Germany. The pressure was calibrated against several metamorphic reactions (Bromiley et al., 2004). Additional hybrid series and serpentinite dehydration series (P27 – P42) as well as mixed series (SB runs) of experiments were carried out at pressures from 0.2 GPa to 0.5 GPa and a temperature of 1250 °C using an internally heated gas pressure vessel at the Korzhinskii Institute of Experimental Mineralogy, Chernogolovka, Russia (Safonov et al., 2014). The duration of hybrid and mixed experiments was from 30 minutes to 120 h (**Supplementary Table A2**). The mixed runs have been performed at controlled temperatures of 1250 - 1300°C and pressures of 0.2 - 0.5 GPa and duration of 0.5, 2, 5 and 48h. During the run quenching down to 550 °C, the pressure was constant, and then slowly released (with average rate of 0.03 GPa/min to 0.08 GPa/min). During the cooling from 1250 to 1000 °C, the cooling rate was 167 °C/min, and then below 550 °C - 90 °C/min.

**Experimental Methods**

Time-dependent experiments were conducted with variable ratios of serpentinite and basalt in the bulk system (from 13.4 to 100 wt% serpentine) and run durations from 1 min to 120 h (**Supplementary Table A2**). Two series of experiments were performed: (a) *hybrid* series with serpentinite cylinder and basaltic powder and (b) *mixed* series composed of intimately and well-mixed serpentinite and basaltic powders. The used $Au_{80}Pd_{20}$ alloy has negligible Fe solubility at redox conditions applied in our experiments (Balta et al., 2011). To avoid any embrittlement of the capsules (related to the Fe pre-saturation) used for the volatile-bearing runs at elevated pressures, the capsules were not pre-doped with Fe. For the hybrid experiments, the experimental design included a serpentinite cylinder (13.4 – 28.2 wt%) in the upper part and basaltic glass powder in the lower part of an $Au_{80}Pd_{20}$ capsule. The mixed runs were performed with mixtures of serpentinite and basaltic powders with 20, 50, 80 and 100 wt% of serpentinite in the mixture (**Figs. 1, 2**). The redox conditions in our experiments were controlled by the initial $Fe^{2+}/Fe^{3+}$ ratios imposed by the starting basaltic glass and serpentinite in the capsule, since a run duration of <48h is too short to reach and maintain equilibrium oxygen fugacity using double-capsule techniques with common mineral buffers (Matjuschkin et al., 2015). The redox conditions established during the runs were estimated from the compositions of co-existing olivine and chromite, using equations (Ballhaus et al., 1991), that yield values from $\Delta QFM$ +1.8 to +5.9 ($\Delta QFM$ denotes log $f_{O2}$ relative to the quartz-fayalite-magnetite buffer, **Supplementary Table A2**). Additionally, the $Fe^{2+}/Fe^{3+}$ ratios in several glass run products were estimated by XANES (X-ray absorption near edge structure) at the European Synchrotron Radiation Facility (ESRF) in Grenoble (France). The oxidized conditions have been established during the runs due to the presence of water and partial $H_2$ loss. The run products (glasses and minerals) were analyzed for chemical and isotopic composition (**Supplementary Tables A3, A4**). To calculate phase proportions, we have used the law of mass conservation and the PYTHON language code. At the first step, the initial MORB glass was introduced instead of $L_{bas}$ and $L_{int}$ (**Supplementary Table A2**). On the second step, the basaltic and interstitial glasses ($L_{bas}$ and $L_{int}$, respectively) were distinguished as two phases. Thermodynamic modelling was performed to constrain the compositions of the fluid released during serpentinite dehydration (**Supplementary Table A5**) and of mineral phases expected to crystallize upon cooling (**Supplementary Tables A6, A7**).

## ANALYTICAL AND MICROANALYTICAL METHODS

### Bulk-rock Analyses for the Starting Materials

The basalt used in the hybrid experiments is a typical Mg-rich mid-ocean ridge basaltic glass (**Supplementary Table A1**) (number 3786/3) from the Knipovich ridge of the Mid Atlantic Ridge sampled by dredging during 38$^{th}$ Research Vessel Adademic Mstislav Keldysh expedition (Sushchevskaya et al., 2000). Serpentinite used as the starting material is homogeneous antigorite-dominated sample with accessory Fe-rich oxides (TSL-19) without any trace of residual chromite, olivine and orthopyroxene from Zildat, the Ladakh area, northwest Himalaya (Deschamps et al., 2010). The serpentinite was prepared as doubly polished ~1000 µm-thick and ~2.6 mm-diameter cylinders. The bulk rock compositions of the serpentinite and basalt samples are given in **Supplementary Table A1** together with those of the reference (UB-N) serpentinite (Govindaraju, 1994). The concentrations of $CO_2$ and S in both rocks were measured at the SARM (Service d'Analyse des Roches et des Minéraux, Centre de Recherches Pétrographiques et Géochimiques, Vandoeuvre lès Nancy, France). The rock $H_2O$ concentrations were measured using Karl Fischer titration. Major and trace elements were measured using inductively coupled plasma optical emission spectroscopy (ICP–OES) and inductively coupled plasma mass spectrometry (ICP–MS), using a method developed at the SARM (Carignan et al., 2001) employing an ICP–OES IRIS Advantage ERS from Thermo Scientific and an ICP–MS x7 from Thermo Scientific.

## Laser Fluorination

Laser fluorination oxygen isotope analyses of serpentinite were performed at the University of Oregon stable isotope laboratory (Bindeman, 2008) using a MAT 253 mass spectrometer integrated to the laser line using home-built airlock sample chamber. Three aliquots of University of Oregon Garnet (UOG) reference material ($\delta^{18}O$ = +6.52 ‰) were analyzed together with the unknown samples during each analytical session and varied ±0.2‰ 1σ standard deviation.

## Scanning Electron Microscope (SEM) and Electron Microprobe analysis and mapping (EPMA)

Major element analyses of minerals and glasses (**Supplementary Tables A3, A4**) and back-scattered electron images of the samples were performed at the Géosciences Environnement Toulouse (GET, Toulouse, France) laboratory and at the Centre de Microcaractérisation Raimond Castaing (Toulouse, France). The main experimental phases in the samples were identified by EDS microprobe technique using a scanning electron microscope (SEM) JEOL JSM-6360 LV with energy-dispersive X-ray spectroscopy (EDS) (Borisova et al., 2013; 2020). Major, minor and volatile element compositions of the crystals and glasses were analyzed using CAMECA SX-Five microprobe and were mapped using CAMECA SX-Five FE at the Centre de Microcaractérisation Raimond Castaing (Toulouse, France), respectively. The X-ray elemental (Si, Al, Ca, Fe, Mg) mapping was performed using CAMECA Five FE. For the mapping of the element distribution, 10 kV and 30 nA with 0.150 s dwell time on the peak and 0.150 s on background were applied for images of 320 to 240 pixels. MgO contents and the other oxide contents were calculated by oxide stoichiometry taking into account the PAP matrix correction mode (Pouchou and Pichoir, 1991). Regularly, electron beam of 15 kV accelerating voltage, and of 20 nA current were applied. Additionally, for the acquisition of minor Cr and Zr elements, we used 200 nA current focused on the sample. We defocused the beam to ~10 µm$^2$ where possible for the hydrous glass analysis with CAMECA SX-Five (Borisova et al., 2020). The following synthetic and natural standards were used for calibration: albite (Na), corundum (Al), wollastonite (Si, Ca), sanidine (K), pyrophanite (Mn, Ti), hematite (Fe), periclase (Mg), Ni metal (Ni), $Cr_2O_3$ (Cr) and reference zircon (Zr). Element and background counting times for most analyzed elements were 10 s (except for 5 s for Na and K) and 5 s, respectively, whereas, peak counting times were 120 s for Cr and 80 to 100 s for Ni and 240 s for Zr. Detection limits for Cr and Ni were 70 ppm and 100 ppm, respectively, and 70 ppm for Zr. The silicate reference materials of MPI-DING series of ultramafic to felsic composition (GOR132-G, GOR128-G, KL2-G, ML3B-G and ATHO-G of Jochum et al., 2006) were analyzed as unknown samples to additionally monitor the analysis accuracy (Borisova et al., 2013). The estimated accuracy ranges from 0.5 to 3 % (1σ RSD = relative standard deviation), depending on the element contents in the reference glasses. Additionally, the silicate reference material analysis allowed to control precision for the major and minor (e.g., Cr, Ni in glasses) element analyses. The estimated precision was in the limit of the CAMECA analytical uncertainty (related to the count statistics).

$H_2O$ contents in the interstitial melts were estimated following the method of Borisova et al. (2005) using the *in situ* analyses of the major element oxides. For this method, we used several reference materials of hydrous glasses with known major element composition and water contents from 0.0, ~4 to ~6 wt% $H_2O$ (Martel et al., 2000; Borisova et al., 2005). The $H_2O$ contents in the glasses obtained using electron microprobe analysis are in good accordance with those of measured independently by secondary ion mass spectrometry.

## Secondary Ion Mass Spectrometry (SIMS)

The $H_2O$ contents and oxygen isotopes in the experimental products were analysed with the CAMECA IMS 1270 E7 ion probe at CRPG-CNRS (Nancy, France), where $H_2O$ and $^{18}O/^{16}O$ ratios were measured simultaneously in multicollection mode. A set of natural (Gurenko et al., 2016) and experimentally produced glasses (Jochum et al., 2006) as well as San Carlos $Fo_{91.3}$ and CI-114 $Fo_{74}$ olivines ($\delta^{18}O$ = 5.2‰, Bindeman, 2008) were used as reference materials to establish instrumental mass fractionation and calibration lines between measured $^{16}O^1H^-/^{16}O^-$ and $^{16}O^1H^-/^{18}O^-$ ratios and reference absolute $H_2O$ concentrations. The external reproducibility measured on the reference glasses was ranged from 0.2 to 8.2% (1σ RSD) depending on $H_2O$ contents in the reference glasses.

$\delta^{18}O_{VSMOW}$ quantification relative to Vienna Standard Mean Ocean Water for olivines has been performed based on San Carlos forsteritic grains analyzed in bracketing mode (Borisova et al., 2016). To calculate oxygen isotope composition of zircon, we applied olivine-quartz-zircon fractionation factors (Chiba et al., 1989; Valley et al., 2003) at the temperature range from 700 to 800°C (**Supplementary Tables A3, A8**).

## Laser Ablation Inductively Coupled Plasma Mass Spectrometry (LA-ICP-MS)

Major and trace element concentrations (**Supplementary Table A3**) were determined by LA-ICP-MS at the Max Planck Institute for Chemistry, Mainz, using a New Wave 213 nm Nd:YAG laser UP 213, which was combined with a sector-field ICP-MS Element2 (Thermo Scientific) (Jochum et al, 2007). The repeatability (RSD) of the measurements is about 1 - 3% (with measurements corresponding to laser crater of 30 µm) and 5 - 10 % (with measurements corresponding to laser crater of 8 µm). The detection limits for the 30 µm measurements vary between about 0.001 and 1 ppm (Jochum et al., 2014). The concentration values agree within about 5 % (30 µm) to 10 % (8 µm) of the reference values (GeoReM database) (Jochum et al., 2005).

## X-ray Absorption Near Edge Structure (XANES) Spectroscopy

Iron redox state in selected quenched glass products was determined from Fe K-edge (~7.1 keV) XANES spectra acquired at the FAME beamline (Proux et al., 2005) of the European Synchrotron Radiation Facility (ESRF). The beamline optics incorporates a Si(220) monochromator with sagittal focusing allowing an energy resolution of ~0.5 eV at Fe K-edge and yielding a flux of >$10^{12}$ photons/s and a beam spot of about 300×200 µm. The samples were finely ground, pressed in pellets of 5 mm diameter and placed in a liquid-helium cryostat (~10 K) to avoid any potential beam-induced degradation of the sample. XANES spectra were acquired in fluorescence mode in the right-angle geometry using a 30-element solid-state germanium detector (Canberra). Energy calibration was achieved using a Fe metal foil whose K-edge energy was set to 7.112 keV as the maximum of the spectrum first derivative. Iron-bearing oxides and silicates with different Fe redox and coordination environment, diluted by mixing with boron nitride to obtain Fe concentrations of a few wt%, were measured similarly to the glasses to serve as reference compounds.

Iron redox state in the glasses was determined by fitting the XANES pre-edge region (7.108-7.120 keV) using the protocols and numerical algorithms developed in Muñoz et al. (2013) for other Fe-bearing minerals, and using a background polynomial and two pseudo-Voight functions to determine the energy position of the pre-edge peak centroid, which is a direct function of the $Fe^{III}/Fe^{II}$ ratio both in crystalline and glass silicate samples (Wilke et al., 2001; 2005). Ferric iron ($Fe^{III}$) mole fraction ($X_{FeIII}$) in the starting basalt, and sample P3 was determined using the calibration established for basaltic glasses (Wilke et al., 2005), while sample P1 that showed a mixture of glass and crystals was processed using the calibration established for Fe minerals (Muñoz et al., 2013). The uncertainties of $X_{FeIII}$ determination for dominantly glassy samples are 0.05 in absolute value, while those for glass-crystal mixtures are typically 0.07 of the value.

## MODELING

### Modelling of Reverse Isobaric Fractional Crystallization

To re-equilibrate the measured compositions of the 0.2 GPa intermediate to felsic melts with the adjacent high-Mg olivine, a modeling of reverse isobaric fractional crystallization has been applied using Petrolog3 software (Danyushevsky and Plechov, 2011). Our model used the conditions of the NNO (nickel - oxide of nickel) redox buffer to constrain $Fe^{II}/Fe^{III}$ in the melts following the model of Borisov and Shapkin (1990). We used the 0.2 GPa pressure condition of the melt water saturation and numerical models of Ariskin et al. (1993) for olivine, clinopyroxene and orthopyroxene, as well as model of Ariskin and Nikolaev (1996) for chrome spinel. The recalculated melt compositions and the corresponding information on the associated minerals are given in the **Supplementary Table A3b**.

## Thermodynamic Modelling of Fluid Composition

Thermodynamic modelling was applied to estimate the composition of the aqueous fluid produced in the experimental serpentinite-basalt systems, assuming local chemical equilibrium within the serpentinite-basalt reaction zone, as supported by the analysis of phase relationships and partition coefficients in the run products (**Supplementary Table A2**). Calculations were performed using the HCh software package and associated Unitherm database, allowing chemical equilibrium predictions in multicomponent fluid-mineral systems based on the minimization of the Gibbs free energy of the system (Shvarov, 2008; 2015), and accounting for non-ideality of the species in aqueous solution using the extended Debye-Huckel equation (Helgeson et al., 1981). The thermodynamic properties of the end members of silicate and oxide mineral phases (e.g., forsterite, fayalite, diopside, enstatite, tremolite, albite, anorthite, sanidine, chlorite, antigorite, chromite, magnetite, corundum) were adopted from Robie and Hemingway (1995) (hereafter RH95 database) and/or from Holland and Powell (2011) (HP11 database), depending on the data availability. Their solid solutions in major silicate minerals such as olivine, pyroxene, amphibole and plagioclase were modelled assuming ideal mixing-on-site models, which is a reasonable approximation for our purposes (Shvarov, 2008; 2015; Holland and Powell, 2011). Furthermore, since both databases report very similar thermodynamic data for the same minerals, the choice of database has no significant effect on the final equilibrium phase composition and abundance. Several silicate glass individual phases equivalent to the corresponding crystalline phase compositions available in the RH95 database ($SiO_2$, $CaMgSi_2O_6$, $CaAlSi_2O_8$, $KAlSi_3O_8$, $NaAlSi_3O_8$, $NaAlSiO_4$) were also used to better account for the presence of melt instead of equivalent crystalline minerals (e.g., feldspars), which were absent in experimental products.

Thermodynamic properties of Na, K, Si, Al, Mg, Ca ions and hydroxide aqueous species described using the revised HKF model (Tanger and Helgeson, 1988; Oelkers et al., 2009) were used from the SUPCRT database (Johnson et al., 1992) with more recent updates for some metal complexes and ion pairs (Shock et al., 1997; Sverjensky et al., 1997; 2014; Tagirov and Schott, 1991; Pokrovski et al., 2013; Zimmer et al., 2016). The HKF model is at present the only one capable of predicting Gibbs energies of many aqueous species and reaction constants at the temperatures and pressures of our experiments (1200 – 1300°C and 0.2 - 1.0 GPa) (Sverjensky et al., 2014; Zimmer et al., 2016). Even though the exact uncertainties of such predictions are strongly element and species dependent and are difficult to quantify rigorously, for the major elements considered above the calculations provide estimates of the order of magnitude of dissolved element concentrations and element relative abundances in the aqueous fluid. Combined with complementary rigorous modelling of fractional crystallization in melt-mineral systems using the rhyolite-MELTS (version 1.0.2) software, the fluid-mineral equilibrium calculations provide a reasonable estimate of the water and dissolved major element amounts that contributed to the generation of interstitial felsic melts.

Calculations were performed at temperature of 1250°C and pressures from 0.2 to 1.0 GPa in a basalt-serpentinite system with basalt-to-serpentinite initial mass ratios ranging from 1:5 to 1:50, as estimated from the extent of the reaction zone in experimental capsules as a function of run duration, temperature, and pressure. **Supplementary Table A5** reports a brief summary of calculation results of dissolved element concentrations in the aqueous fluid phase for the basalt to serpentinite ratios of 5:1, 10:1 and 50:1 at 1250°C and 0.2, 0.5 and 1.0 GPa. The major mineral phases predicted to be formed at equilibrium are olivine and orthopyroxene, together with minor amounts of clinopyroxene, magnetite, chromite and feldspar-like glass phases. This solid phase composition is in good agreement with that observed in most experiments. The calculations predict forsteritic olivine ($Fo_{93-96}$ mol%) and orthopyroxene (Mg# 94-96) compositions within the range analyzed in the quench products (**Supplementary Table A3, A4**), suggesting local thermodynamic equilibrium between the fluid and these minerals. The aqueous fluid produced by breakdown of serpentinite (in the presence of basaltic melt) is enriched in Si (up to 5 wt%), Na (up to 2 wt%), Al (up to 0.2 wt%) and K (up to 0.4 wt%), with other elements being much less abundant (Mg, Ca, Fe < 1.0-0.1 ppm). Pressure increase from 0.2 to 1.0 GPa generally results in an increase of all element concentrations in the fluid, in agreement with the increase of water solvent density and water solvation power (Pokrovski et al., 2013). A typical order of decreasing concentration in the fluid is the following: Si>Na≈Al>K>>Fe>Ca≈Mg. The much lower concentrations of Fe, Ca, and Mg may, at least partly, reflect the paucity of data for their aqueous hydroxide complexes at elevated temperatures and pressures. Furthermore, the available thermodynamic data do not include metal-silicate complexes known for Al, Fe, Ca and Mg in lower-temperature fluids (Salvi et al., 1998; Pokrovski et al., 2002) and ignore the potential presence of salt (chloride ligand) in natural hydrothermal fluids. If such silicate and chloride complexes are stable at the conditions of our experiments and those of the Hadean lithosphere, the aqueous contents of these metals are expected to increase, but the general order of element enrichment is unlikely to change significantly. Overall, the modelling results demonstrate the high solubilities of the key elements, Si, Na, Al and K in the aqueous fluid phase, sufficient to account for the element contents and ratios in the formed interstitial glass of compositions close to that of the continental crust (**Figs. 2, 3**).

## Thermodynamic Modelling of the Felsic Liquid Crystallization

Thermodynamic equilibrium modelling was applied to obtain information on mineral and fluid phases appearing during the fractional crystallization of the felsic melts on isobaric cooling at 0.2 GPa and to simulate the further evolution of the formed felsic melts in lithosphere settings. We performed the thermodynamic modeling because such data cannot be accessed from our duration-limited laboratory experiments. The modelling was performed using the rhyolite-MELTS (version 1.0.2) software (Ghiorso, and Sack, 1995). Modelling of the mineral crystallization is presented in the **Supplementary Table A6**.

## Thermodynamic Modelling of Zircon Crystallization in the Felsic Melt

Modelling of zircon crystallization is presented in the **Supplementary Table A7**. The primary thermodynamic condition for zircon nucleation and growth from a cooling felsic melt is the melt supersaturation with zircon during a cooling history ($C_{sat} \leq C_{melt}$) (Bindeman and Melnik, 2016). Following Boehnke et al. (2013), zircon saturation $C_{sat}$ in felsic melts is a function of temperature $T$ (in Kelvin) and compositional factor M = (Na+K+2Ca)/(Al×Si), where the elements are given in molar quantities:

$$C_{sat} = 500000 / \exp\left(\frac{10108}{T} - 1.16(M-1) - 1.48\right) \qquad \text{(eqn. 1)}.$$

Values of $C_{sat}$ are presented in **Supplementary Table A7**. For the measured melts initially containing 70 - 120 ppm Zr (120 ppm is maximal Zr content), zircon growth may be initiated at temperatures below 750 ºC for all values of M. We also implemented a linear dependence of M factor on temperature following Harrison et al. (2007) at each computational step:

$$M = 0.5 + 0.0013\ T(°C) \tag{eqn. 2}$$

Similarly, equation 2 suggests that crystallization of zircon from felsic melt with 120 ppm Zr and *M* = 1.5 will start at temperatures below 750 °C. If the felsic melt contains 70 ppm Zr (as the average Zr content in the measured P33 glasses), and M factor ranges from 1.4 to 3.5 (for the case of P32, P33, P35, P37 and P42 glasses), zircon will start to crystallize at even lower temperatures.

Additionally, the zircon saturation $C_{sat}$ in felsic melts was re-evaluated using equation 3 according the Watson and Harrison (1983):

$$\ln D_{zrc/melt} = -3.80 - (0.85(M-1)) + \frac{12900}{T} \tag{eqn. 3},$$

where $D_{zrc/melt}$ is equilibrium zircon-melt partitioning, M is again calculated according the eqn. 2 and T is absolute temperature in Kelvin. Overall, the produced felsic melts may crystallize zircon below 750°C.

# RESULTS

## Experimental Sample Description

In this section, we describe twenty three experimental runs on (*i*) pure serpentinite dehydration and (*ii*) serpentinite-basalt interaction in the mixed and hybrid systems with different duration and different proportion of serpentinite (13 – 100 wt%) in the starting material performed at 0.2 to 1.0 GPa and 1250 – 1300°C (**Supplementary Table A2**).

## Experiments on Serpentinite Dehydration and Mixed Serpentinite-Basalt Interaction at 0.2 GPa Pressure

Three experiments on pure serpentinite dehydration (100 wt% of serpentinite) at 0.2 GPa and 1250° C were performed with different run durations. In case of the short-time experiment (P29), we observed a heavily porous aggregate consisting of enstatite, olivine, and chromite (**Supplementary Tables A2 and A3**). For a long-duration experiment P28, the experimental products were represented by crystallized powder of the same mineralogical composition. No trace of silicate glass was found in the runs modelling serpentinite dehydration.

Based on textural and compositional characteristics, three types of mineral-glass assemblage have been distinguished in nine samples of the mixed experiments on serpentinite-basaltic melt interactions with different proportion of serpentinite (20 – 80 wt%) in the starting material. (a) **Figure 1** demonstrates that the samples SB7, SBbis2, SBter3 with highest initial proportion of serpentinite (80 wt%) are characterized by unique polyhedral olivine-rich zone of unzoned crystals of olivine associated with interstitial felsic glasses ($L_{int}$) and interstitial crystals of clinopyroxene and fluid bubbles. Chromite is associated with the euhedral olivine grains (forsteritic olivine $Fo_{90-94}$). (b) Samples number SB1, SB4, SBbis1, SBter2 with lower initial proportion of serpentinite (20 – 50 wt%) are represented by polyhedral olivine phenocrystals. In the matrix, these samples contain clinopyroxene-olivine assemblage with interstitial felsic glasses ($L_{int}$). Oxide minerals are represented by chromite phenocrysts. (c) Finally, the sample SBter1 with lowest initial proportion of serpentinite (20 wt%) is represented by homogeneous basaltic glass likely formed by complete hybridation of the starting basaltic liquid with serpentinite, whereas the sample SBbis3 with the same proportion of serpentinite (20 wt%) is characterized by residual olivine-rich zone. In general, most samples of the SB series with high initial proportion of serpentinite (50 – 80 wt%) are characterized by homogeneous polyhedral olivine-rich zone of unzoned crystals of forsteritic olivine ($Fo_{90-94}$ mol%) associated with interstitial felsic glasses ($SiO_2$ = 62 – 71 wt%) and crystals of clinopyroxene (Mg# 58 - 68), as well as fluid bubbles (**Fig. 1**). Chromite (Cr# = 68 – 89; Mg# 56 - 73) is associated with the euhedral olivine grains (forsteritic olivine $Fo_{90-94}$ mol%).

**Hybrid Experiments on Serpentinite-Basalt Interaction at 0.2 GPa**

The hybrid samples with low initial proportion of serpentinite (18 – 28 wt%) in the starting material produced at 0.2 GPa pressure have two distinct zones: former-serpentinite (or olivine-rich) zone and quenched basaltic melt zone (**Fig. 1; Supplementary Tables A2 and A3**). The olivine-rich zone in the sample P37 is represented by enstatite (Mg# = 96.5 ± 2.1) with olivine ($Fo_{93±1}$), and olivine-rich periphery (from 80 to 280 μm thickness) consisting of olivine and interstitial felsic glass. Sample P32 was maintained during 2.5 hours at 1250° C and 0.2 GPa. The former-serpentinite zone has an inner harzburgite part and an outer olivine-rich part (olivine Fo$_{89\pm1}$ and orthopyroxene Mg# = 95 ± 3) composed of olivine and interstitial glass. Chromite is localized exclusively in the former-serpentinite zone (**Fig. 1**). P35 sample ran for 5 hours at 1250° C and 0.2 GPa. In the former-serpentinite zone at the point of contact with hydrous basalt glass, a large (~200 × 290 μm) aggregate of chromite has been discovered. Gradually moving away from the contact to the former-serpentinite zone center, pockets of interstitial glass with rare crystals of amphibole, zoned olivine (more magnesian in the core and more ferric at the rim) with less amount of interstitial glass in the olivine-rich area consisting of olivine and enstatite (olivine Fo$_{80\pm7}$ and orthopyroxene Mg# = 92 ± 2) were observed. The run P33 was performed at 0.2 GPa and 1250°C during 48 hours. The olivine-rich zone comprises euhedral zoned crystals of olivine with highly magnesian cores and more ferrous rims associated with interstitial felsic glasses and interstitial crystals of clinopyroxene, amphibole and fluid bubbles (**Fig. 1**). Chromite and Cr-bearing magnetite are located in the interstitial zones between the euhedral olivine grains (olivine Fo$_{90\pm4}$) and are also associated with amphibole microcrystals grown around the Cr-rich spinels. Only one crystal of ferrous orthopyroxene (hypersthene) was found in association with interstitial glass. P42 sample is represented by quenched melt zone of hydrous basaltic glass with rare euhedral olivine. The olivine-rich zone is represented by euhedral forsteritic crystals (Fo$_{92.8\pm0.5}$), chromiferous magnetite and interstitial felsic glasses of basaltic andesite composition.

**Hybrid Experiments on Serpentinite-Basalt Interaction at 0.5 GPa**

The following group of samples with low initial proportion of serpentinite (13 – 19 wt%) in the starting material was produced at pressure of 0.5 GPa and temperature of 1300 °C (except P36 with temperature 1250°C) with duration of the experiment from 1 minute to 8 hours (**Supplementary Tables A2 and A4**). In the sample P1 with shortest duration of the experiment (1 minute), the quenched glass zone and former-serpentinite zone are present. The former-serpentinite zone contains fine-grained (5 to 10 µm in size) aggregates of olivine $Fo_{95}$, enstatite with Mg# = 95, chromite (Cr# = 89) and interstitial glass of basaltic andesite composition. Chromite crystals (first microns in size) are disseminated within this zone. The bubbles in the quenched melt zone are evidence for water fluid presence during quenching. Sample P15 was maintained during 0.5 hours at 0.5 GPa and 1300 °C. Two distinct zones (quenched basaltic melt zone and olivine-rich zone) are also present in this sample. The quenched basaltic melt zone consists of hydrous basaltic glass characterized by high magnesium content (11 wt.% MgO). The former serpentinite zone consists of 5-20 µm-size aggregate of olivine $Fo_{93}$ and enstatite with Mg# = 97. It is associated with interstitial glass of andesitic composition and chromiferous magnetite. The outer part of this zone is represented by a layer (~120 µm thickness), consisting of olivine crystals and interstitial glass. Chromiferous magnetite is concentrated in two large areas (up to 200 µm and 400 µm in size each) or disseminated near these areas. Sample P18 provides information about 5 hours serpentinite-basalt interaction. The sample is represented by the hydrous basaltic zone and former-serpentinite zone (**Supplementary Tables A2 and A4**). The olivine-rich zone consists of two areas. The outer area of nearly 200 µm width contains olivine and basaltic interstitial glass with disseminated chromite (first microns in size). The inner area is situated 200 µm from quenched basaltic glass. It consists of olivine $Fo_{93}$, enstatite with Mg# = 97 and contains aggregate of magnetite (~20 × 40 µm). P36 is the longest (8 hours) experiment of this series at 0.5 GPa. Serpentinite was completely dissolved in the basalt melt (**Supplementary Tables A2 and A4**). Run products show homogeneous basaltic glass with high MgO content (11.5 wt%). Numerous bubbles with dendritic amphibole aggregates are present in the glass.

**Hybrid Experiments on Serpentinite-Basalt Interaction at 1.0 GPa**

Two experiments with low initial proportion of serpentinite (15 – 16 wt%) in the starting material were performed at 1.0 GPa and 1300 °C. Experiment P3 with duration 2.5 hours contains basaltic glass zone and former-serpentinite zone (**Supplementary Tables A2 and A4**). Forsteritic olivine (Fo$_{92}$), enstatite with Mg# = 96 and Cr-bearing magnetite are associated with interstitial glass of basaltic composition (up to 51 wt% SiO$_2$) in the former-serpentinite zone. The 9 h-long run (P7) contains hydrous basaltic glass with 13.0 wt% of MgO (**Supplementary Tables A2 and A4**).

**Synthesis of the Experimental Data**

Complete dissolution of dehydrated serpentinite in the basaltic melt and production of homogeneous crystal-free hydrous basaltic melts (L$_{bas}$) were observed in hybrid and mixed experiments with lowest proportion of serpentinite (≤ 20 wt%) in the starting material and longer than 5 h at pressures of 0.2 – 1.0 GPa (**Fig. 2; Supplementary Tables A2 - A4**). In the shortest (<5h) hybrid runs at pressures of 0.5 – 1.0 GPa, and in all hybrid runs at 0.2 GPa lasting 0.5 to 120 h, two zones were found. One zone is composed of hydrous basaltic glass and the other zone consists of olivine and interstitial glass (L$_{int}$) (former-serpentinite zone or olivine-rich zone) (**Figs. 1, 2, Supplementary Tables A3, A4**). In the runs on serpentinite dehydration containing 100% serpentinite as starting material, highly magnesian olivine (Fo$_{91-92}$, mol% of forsterite molecule), orthopyroxene (92 – 95 Mg# = 100 Mg/(Mg+Fe$^{2+}$)) and chromite (69 – 73 Mg#, and 68 – 95 Cr# = 100 Cr/(Cr + Al)) were found, but no interstitial silicate glass was detected. In contrast, a homogeneous olivine-rich zone containing mostly olivine (Fo$_{90-94}$, mol%), chromite (56 – 73 Mg#, and 68 – 89 Cr#) and interstitial glass was observed in all mixed experiments at 0.2 GPa containing the highest proportion (50 – 80 wt%) of serpentinite in the starting material. Thus, the analyzed olivine-rich zones contain chromite (36 – 73 Mg#, and 68 – 89 Cr#), magnesian olivine (Fo$_{80-94}$ mol%) and, in the shortest hybrid runs (≤ 5h), magnesian orthopyroxene (92 – 95 Mg#) in association with interstitial glass. In most experiments at pressures of 0.2 to 0.5 GPa, fluid bubbles are present, presumably produced by serpentinite decomposition. Olivine in the 0.2 GPa experiments is enriched in $^{18}$O ($\delta^{18}$O$_{VSMOW}$ = +6.2 to +6.6‰), in accordance with $^{18}$O enrichment of the starting serpentinite ($\delta^{18}$O$_{VSMOW}$ = +9.7 ‰, **Supplementary Tables A1, A3**) relative to the typical mantle value of +5.3 ‰ (Hoefs, 2005).

**Interstitial Glass Composition**

In the hybrid and mixed runs at 0.2 to 1.0 GPa, the olivine-rich zones host interstitial glass pockets of 10 to 200 µm in size. The glass is enriched in silica, aluminum, alkalis and water contents (**Fig. 3, Supplementary Tables A2 - A4**) relative to the starting basaltic material. The major element composition of the glasses produced in those experiments is mainly controlled by pressure, but also depends on the basalt/serpentinite proportion in the mixed runs. The glasses produced in the hybrid run lasting 2.5 h at 1.0 GPa are mafic (50 wt% SiO$_2$) with an average MgO content of 10 wt%, similar to those produced in previous studies of equilibrium partial melting of hydrous peridotite at similar pressures (Ulmer, 2001). In contrast, all hybrid runs at 0.2 GPa, and the mixed experiments at 0.2 GPa with the highest proportion of serpentinite (50 – 80 wt%), as well as short (≤ 0.5 h) runs at 0.5 GPa produced glasses showing generally higher SiO$_2$ (57 to 71 wt%) contents. The both measured and the recalculated glass compositions vary from basaltic andesite to dacite and are consistent with the general tendency of olivine-saturated partial melts to become progressively richer in $SiO_2$ with decreasing pressure and higher water contents (Hirschmann et al., 1998). For example, the mixed runs with highest proportion of serpentinite (80 wt%) produced dacite glasses up to 62 – 63 wt% $SiO_2$ contents in the recalculated composition (**Figs. 3, 4, Supplementary Table A3**). Furthermore, the 0.2 GPa recalculated glasses are enriched in alkalis (up to 3.9 wt% $Na_2O + K_2O$), aluminum (up to 16.7 wt% $Al_2O_3$) and chromium (up to 1400 Cr ppm) (**Figs. 3, 4; Supplementary Tables A3, A4**). Most glasses produced at 0.2 GPa are more enriched in Si, Al, and alkalis compared to those of dry glasses produced at 0.1 MPa (Fisk, 1986) as well as nearly dry glasses produced at 0.5 GPa (Kelemen et al., 1995).

The intermediate to felsic glasses were also analyzed for trace element contents. A strong positive Pb anomaly relative to light rare earth elements (LREE) observed in the felsic glasses is an important signature of typical continental rocks compared to the mantle-derived magmas (Rudnick and Gao, 2003). The incompatible trace element patterns of the felsic glasses are nearly identical to those of the modern TTG magmas found in the mantle sections of ophiolites (Amri et al., 1996; 2007; Shervais et al., 2008; Xu et al., 2017). The felsic glasses produced in runs at 0.2 GPa have major (Si, Ti, Al, Mg, Ca, Na, K, P) and trace (Cr, Sr/Y, Th/U and $Ce_N/Yb_N$, where N denotes primitive mantle-normalized (Sun, and McDonough, 1998; Lyubetskaya and Korenaga, 2017) element contents and ratios, as well as Mg/(Mg+Fe), that are also similar to those of the modern TTG situated in the mantle sections of ophiolites. The 0.2 GPa intermediate to felsic melts are characterized by lower $SiO_2$ and $Na_2O + K_2O$ contents and higher Cr contents in comparison to those of Archaean TTG rocks. The slight enrichment in LREE of the felsic glasses differs from strongly fractionated patterns of the Archaean tonalite-trondhjemite-granodiorite (TTG) magmas (Moyen and Martin, 2012). This is in line with the initial conditions of the shallow felsic crust formation by a mechanism distinct from plate subduction operating during the Archaean eon (Moyen and Martin, 2012). Except for fluid-mobile B and Pb, trace element compositions of the 0.2 GPa interstitial felsic glasses are similar to those of the starting basaltic melt (**Fig. 5**). Additionally, hafnium contents (0.5 – 4.0 ppm Hf) and Lu/Hf ratios ($Lu_N/Hf_N$=0.3 – 2.2) in the 0.2 GPa felsic melts are similar to those of the starting basaltic melt (2.7 ppm and 0.7, respectively), suggesting that the experimental felsic melts produced at 0.2 GPa might have inherited their Hf isotope signatures from the precursor basaltic reservoir (**Supplementary Table A3**).

**Results of Thermodynamic Modelling**

Thermodynamic modelling shows that the aqueous fluid generated by serpentinite dehydration in the presence of basaltic melt is enriched in Si, Al, Na and K (**Supplementary Table A5**) compared to both starting and reacted serpentinite. The major mineral phases predicted to be formed at equilibrium are olivine and orthopyroxene, as well as minor amounts of clinopyroxene, magnetite, chromite and felsic melt phases. This assemblage is in good agreement with that observed in most hybrid and mixed experiments (**Supplementary Tables A2 - A4**). Calculations predict olivine ($Fo_{93-96}$ mol%) and orthopyroxene (Mg# 94-96) compositions within the range of those analyzed in the minerals and quenched glasses, suggesting local thermodynamic equilibrium between the fluid and the produced minerals. The aqueous fluid produced by breakdown of serpentinite, in the presence of basaltic melt, is enriched in Si (up to 5 wt%), Na (up to 2 wt%), Al (up to 0.2 wt%) and K (up to 0.4 wt%), with other elements being much less abundant (Mg, Ca, Fe < 1.0 - 0.1 ppm). This result emphasizes the role of the basaltic component that provides key fluxing elements to the bulk hybrid system and the aqueous fluid and drives generation of the felsic melts. Indeed, the textures on the **Figure 1** demonstrate that formation of the intermediate to felsic liquids preceded the complete dissolution of orthopyroxene in the olivine-rich zones of the shortest experiments (e.g., P1 run).

The hydrous intermediate to felsic melts produced in our experiments were likely formed by the fertilization of olivine-rich zones due to the presence of the aqueous fluid owing to hydrated peridotite interaction with basaltic melt at 0.2 – 1.0 GPa. This process is coupled with incongruent melting of harzburgite producing olivine at the expense of orthopyroxene, mostly at 0.2 GPa. The presence of polyhedral, mostly unzoned olivine crystals, co-existing with homogeneous interstitial felsic melt in the olivine-rich zone of the longest runs (**Fig. 1**) (**Supplementary Table A2**) supports new olivine growth and possible local olivine-melt equilibrium at 1250°C and 0.2 GPa. However, intra- and inter-sample variability of the measured glass composition implies a lack of equilibrium between the interstitial melt and the host olivine and that the serpentinite source was not completely homogeneous in the portions used for the experiments. Additionally, the derived Fe-Mg partition coefficients between olivine and the associated felsic melt are low (**Supplementary Table A2**), indicating a lack of equilibrium. The hydrous andesitic melts with ~5 wt% $H_2O$ have very low the glass transition temperatures of Tg ≈ 300 °C (e.g. Deubener et al., 2003), and thus these melts have particular susceptibility to being quench modified. Reverse fractional crystallization calculations of the measured glasses to attain thermodynamic equilibrium with the associated high-Mg olivine require the fractional crystallization of olivine ± clinopyroxene + chromite phases, which are observed among the quench and olivine rim products. These calculations require the recalculated interstitial melts – olivine equilibrium temperature ranging from 1150 to 1312°C (**Supplementary Table A3**) and approaching the run temperature of 1250°C. These calculations indicate isobaric fractional crystallization of olivine (up to 18%) ± clinopyroxene

+ chromite (up to 0.3%) phases upon quenching. Nevertheless, the recalculated melts have intermediate to felsic composition with strongly elevated Mg index (**Fig. 4**).

Our thermodynamic simulations of the intermediate to felsic melt crystallization using the rhyolite-MELTS (version 1.0.2) software predict that upon cooling from 980 to 700 °C in the presence of an aqueous fluid, the felsic melt would predominantly crystallize quartz and feldspar, but also amphibole and biotite (**Supplementary Table A6**). Such felsic magmas share similarities with the TTG intrusions in mantle peridotites of ophiolites (Amri et al., 1996; 2007; Shervais et al., 2008; Xu et al., 2017). Thermodynamic modelling also shows that our felsic melts would precipitate zircon upon cooling below ~750 °C (**Supplementary Table A7**), which agrees with findings of zircons in TTG intrusions of the modern ophiolites (Rioux et al., 2013).

## DISCUSSION

## Shallow Intermediate to Felsic Crust Production by Melting of Hydrated Peridotite

Our study brings new constraints on the mechanisms of intermediate to felsic crust production. Hydrated peridotites and mafic cumulates are the most abundant constituents of the modern slow-spreading oceanic lithosphere at the Moho mantle-crust transition zone level (Cannat, 1993). Generation of an intermediate to felsic crust within the peridotite lithosphere of the modern Earth at high temperature and shallow (< 10 km) depths likely result from cyclic magmatic and hydrothermal processes. These include repetitive intrusions of mantle-derived basaltic magmas below spreading centers (Amri et al., 1996; 2007; Borisova et al., 2012), where reaction between high-temperature basaltic melts and a previously formed cold hydrated ultramafic lithosphere is possible during periods of low magma supply and oceanic water circulation at shallow depth (**Fig. 6**). The hydrated peridotite melting produced in our experiments created dunite- or chromitite-type rocks (**Figs. 1, 2**) similar to their natural analogs described in the ophiolites (e.g., Dick, 1977; Benoit et al., 1999; Borisova et al., 2012; Zagrtdenov et al., 2018; Rospabé et al., 2019). Our experimental data thus confirm an empirical model of the hydrated mantle-basaltic magma reaction inferred from natural data on chromite-hosted inclusions from the Oman ophiolite (Borisova et al., 2012). The olivine-rich system containing interstitial hydrous quartz-saturated liquids, like those obtained in our experiments, is gravitationally instable because of the high density contrast between hydrous felsic melts (<2.7 g/cm$^3$) and mafic minerals (3.2 - 3.3 g/cm$^3$) and chromite (~5.0 g/cm$^3$). Referring to processes happening at spreading centers (Amri et al., 1996; 2007; Benoit et al., 1999), and that can be observed in ophiolites, the low-density hydrous felsic melts generated below and/or at the Moho transition zone are collected in pods and veins tens of centimeters to meters in size and can progressively coalesce into larger intrusions in the shallow oceanic lithosphere. Alternatively, the felsic melts produced by aqueous fluid-assisted partial melting of peridotite may remain scattered in peridotite at shallow depths <10 km and may be sampled as olivine-hosted melt inclusions in peridotite xenoliths (Hirschmann et al., 1998). Upon segregation and cooling owing to the hydrothermal circulation, these hydrous felsic liquids will crystallize as assemblages enriched in quartz and Ca-rich plagioclase (up to An$_{95}$) similar to those recorded in the mantle and lower crustal sections of the Oman ophiolite (Amri et al., 1996; 2007) (**Supplementary Table A6**). In addition to low-temperature plagiogranite magmas formed in the oceanic crust due to partial melting of hydrated gabbros (Koepke et al., 2004), high-temperature plagiogranites are formed through aqueous fluid-assisted partial melting of peridotite.

## Implications for the Early Earth and Production of TTGs

Hydrated peridotite, and its chemical analog serpentinite, was likely to be a major shallow component of the Hadean and Noachian protocrusts on Earth and Mars (Albarede and Blichert-Toft, 2007; Elkins-Tanton, 2012), formed by interaction of seawater-derived fluids with peridotites (Guillot and Hattori, 2013). Indeed, planets like Earth and Mars were massive enough to make possible the formation of an early ultramafic (peridotite-like, silica-poor) magma ocean, whereas the planet's distance from the Sun and the surface temperature were appropriate for the existence of an ocean of liquid water (Valley et al., 2002; Albarede and Blichert-Toft, 2007; Müntener, 2010). It is widely recognized (Albarede and Blichert-Toft, 2007; Elkins-Tanton, 2012) that magma ocean(s) underwent solidification to produce shallow ultramafic-mafic protocrust at the earliest stages of planetary evolution, although it remains unclear how such rocks could generate early Hadean intermediate to felsic, quartz-normative to quartz-saturated magmas. Direct natural evidence for the precursor of Hadean ($\geq 4$ Ga) felsic crust is scarce, and remains the subject of intense debate (Wedepohl, 1995; Rudnick and Gao, 2003; Harrison, 2009; Reimink et al., 2014; 2016; Burnham and Berry, 2017; Bell et al., 2018). Investigations of the 4.37 – 4.02 Ga Jack Hills detrital zircons (JHZ), which are the oldest remnants of the primordial felsic magmas on Earth, support the existence of a felsic crust at the earliest stages of planetary evolution in the Hadean eon (Cavosie et al., 2006; Harrison, 2009; Burnham and Berry, 2017). Either reworking into younger crust or recycling of the Hadean crust into the mantle has been hypothesized to explain the lack of remnants of the earliest felsic crust on Earth (O'Neil and Carlson, 2017). However, studies of the Jack Hills zircons are now used to infer the presence of an igneous source that had experienced low-to-moderate temperature aqueous alteration, thereby ruling out a sedimentary source (Burnham and Berry, 2017; Whitehouse et al., 2017), although alternative opinion still exists (Harrison, 2009; Bell et al., 2018). Thermodynamic modelling performed here shows that the 0.2 GPa felsic melts have zircon on their liquidus, upon cooling to temperatures below 750 °C (**Supplementary Table A7**), in agreement with temperature estimates for the Hadean detrital zircons found in the Jack Hills metaconglomerates (Cavosie et al., 2006; Harrison, 2009; Burnham and Berry, 2017), which demonstrate the predominance of hydrous granitic minima melts. The calculated $^{18}$O enrichment in zircons produced from the 0.2 GPa interstitial felsic melts ($\delta^{18}O_{VSMOW}$ = +7.1 to +7.7 ‰, **Supplementary Table A3**) may happen if the parental serpentinite reservoir, formed at temperature below 100 – 125°C, that shifts $\delta^{18}O_{VSMOW}$ above 5.3‰ mantle values (Savin and Lee, 1988), has been rapidly dehydrated at high temperature. For example, the dehydration may be triggered by basaltic melt intrusions. In this way, the aqueous fluids liberated during dehydration may interact with basaltic melt to produce higher than the 5.3‰ mantle in $\delta^{18}O$ felsic melts crystallizing zircon, which will inherit this $^{18}$O-enriched signatures and preserve them in the geologic record (**Supplementary Table A8**). Available data show higher than the mantle $\delta^{18}O$ values (Wilde et al. 2001) but also significant $\delta^{18}O$ heterogeneity of the Jack Hills Zircon grains. For example, recent analyses of concordia age-constrained grains do show $^{18}$O enrichment (7.8 ± 0.4 ‰, Valley et al., 2002; Cavosie et al., 2006; Whitehouse et al., 2017) which is comparable to that estimated in our study. Concerning Hf signatures in the JHZ, the filtered concordia age-constrained data are not sufficient to confirm their 'mafic protocrust' origin (Whitehouse et al., 2017). To explore this idea further we have predicted the trace element composition of melts in equilibrium with Jack Hills zircons (JHZ), based on the zircon composition (Cavosie et al., 2006) and zircon-melt partitioning (Sano et al., 2002; Luo and Ayers, 2009). The primitive felsic melts obtained in our experiments at 0.2 GPa have low $P_2O_5$ contents (≤0.16 wt% for the recalculated glasses), and $Ce_N/Yb_N$ ratios (1.7 – 7.4 for the measured glasses). Such compositions are comparable to those of ancient JHZ melts, which were predicted to have $P_2O_5 \leq 0.1$ and $Ce_N/Yb_N = 0.4 - 18$ (**Figs. 3, 4, Supplementary Table A9**). Furthermore, experimental felsic melts have Th/U ratios (0.8 – 9.4) and Y contents (18 – 29 ppm), which compare favorably with those of theoretical melts in equilibrium with JHZ at 800°C (Th/U = 0.4 – 1.0 ppm and Y = 1.7 – 26 ppm in the JHZ magmas). In addition, Th/U ratio decrease and strongly incompatible LREE concentrations increase as felsic magmas evolve through fractional crystallization (**Fig. 3**). Thus, there are strong chemical similarities between the experimental felsic melts produced at 0.2 GPa with the highest serpentinite fraction (~80 wt%) studied in our work and the JHZ melts. It should be noted that the Hadean peridotite and basalts may be much less depleted in incompatible elements than their modern analogs. Indeed, the fact that the JHZ melts are enriched in most incompatible elements compared to those of the 0.2 GPa intermediate to felsic melts produced in our experiments, performed using modern-day lithologies (**Fig. 5**) may be explained by more depleted modern-day compositions. This incompatible element signature aside, we note that the JHZ fragments contain inclusions of quartz, biotite, apatite, amphibole and feldspar (Harrison et al., 2009; Bell et al., 2018), most of which are shown to be thermodynamically stable in our simulations of the experimental felsic melt crystallization upon cooling (**Supplementary Table A6**). This result is consistent with a magmatic origin for the mineral inclusions, although some part of these inclusions may have a secondary metamorphic origin (Bell et al., 2018). Thus, intermediate to felsic liquids (up to 66 – 71 wt% $SiO_2$) experimentally produced at 0.2 GPa are possible parental melts to the Jack Hills zircons at 4.37 – 4.02 Ga. In contrast, at similar $SiO_2$ contents, higher Cr contents (up to 1400 Cr ppm) in the experimental felsic glasses produced at 0.2 GPa differ strongly from those of the Acasta Gneiss Complex magmas (4.02 – 2.94 Ga), suggesting no genetic relationship with these types of magmas.

We here propose that during the earliest stages of Earth evolution, intrusions of heat-pipe basaltic melts into ultramafic-mafic protocrust might have initiated the dehydration process. The released fluid promoted fertilization and partial melting of the peridotite, whereas the partial melting produced tonalite-granodiorite magmas which formed intrusion bodies in the peridotite protocrust at shallow <10 km depths (**Fig. 7**). The low density of the hydrous melts and the lithosphere hydraulic fracturing were likely the main factors favorable for felsic melt segregation to the upper protocrust. Alternatively, generation of such high-temperature intermediate to felsic melts and magmas within the early ultramafic-mafic protocrust at shallow depths may have also been possible due to impact-induced protocrust melting (Marchi et al., 2014) at conditions of the predominant proportion of serpentinite above 50 wt% into the hybrid system. Such melting would have favored mixing of shallow hydrated ultramafic and mafic components at high temperature well above the liquidus of basalt before the onset of plate tectonics (**Fig. 8**).

## Implications for the Early Mars

The proposed mechanism of aqueous fluid-assisted partial melting of peridotite may have led to formation of the shallow felsic (continental) crust on other rocky planets such as Mars, provided either liquid water or volatile element-rich magma ocean was present at early stage of their evolution. The finding of ancient serpentinites at the surface of Mars (Ehlmann et al., 2010) suggests that serpentinite formation and aqueous fluid-assisted melting of hydrated peridotite may be envisioned during the earliest stages of the Solar System evolution. There is also evidence for the 4.43 – 4.13 Ga-old granodiorite continental crust on early Mars (Sautter et al., 2015; 2016). These quartz-normative rocks, which are believed to represent the early

Martian continental crust, cannot be produced from partial melting of the mantle and subsequent fractional crystallization (Sautter et al., 2015; 2016). By contrast, recent modeling using the MELTS software (Udry et al., 2018) suggest a possibility that the Gale crater felsic rocks were produced by accumulation/fractionation of feldspar from basaltic melt; however, no comparison with the modern TTG rocks was made in that study. Interestingly, the early Noachian granodiorite rocks reported on Mars (Sautter et al., 2015; 2016) have Si, Al, alkalis and the other major element contents (**Figs. 3, 9**) close to those of the modern TTG from the mantle sections of Earth ophiolites (except for the lower 100 Mg/(Mg+Fe) index of peridotite-derived magmas originated from Fe-rich Martian mantle) (Sautter et al., 2015; 2016; Udry et al., 2018). Thus, these data are also in line with our model and suggest that the Noachian granodiorite rocks on Mars may have been produced by aqueous fluid-assisted partial melting of peridotite induced by reaction with basaltic melts.

The process of magma ocean solidification on Mars might have been favorable for the production of the silica-enriched melts (Bouvier et al., 2018). The primordial magmatic environment of peridotite in the presence of silicate melt at shallow depth is comparable to the partially melted volatile element-enriched magma ocean(s) of ultramafic-mafic composition. Such environments are favorable for production of silica-enriched melts in the presence of elevated water contents. The mechanism suggested in our work may thus be responsible for the formation of the earliest crust. Survival of the primary intermediate to felsic liquids and formation of a solid crust may have been possible at conditions of a limited effect of gravitational overturn (Elkins-Tanton, 2012; Bouvier et al., 2018).

# CONCLUSIONS

In summary, the proposed mechanism of aqueous fluid-assisted partial melting of peridotite induced by reaction with basaltic melt supports the possibility of the planetary felsic crust formation at depths of ≤10 km. Such shallow depths of generation of felsic magma revealed in this study implies that formation of silica-rich crust does not require convergent plate tectonics, in agreement with geodynamic simulations of the early Earth's and Mars's hotter mantle (Herzberg et al., 2010; Sautter et al., 2016). Such conditions for the high-temperature reactions between hydrated peridotite and basaltic melts were highly probable in spreading centers, hot spots and Hadean heat-pipe volcanoes. The felsic crust may also have been formed in a more transient way during intensive impact-induced melting of the ultramafic-mafic protocrust in the presence of liquid water ocean on early Earth and Mars. Even though the geodynamic and thermal conditions on both planets have changed since then, felsic crust is still forming in small volume at high temperature of spreading centers, where it essentially stays scattered in the oceanic lithospheric mantle on modern Earth.

# ACKNOWLEDGEMENTS

This work is dedicated to the memory of E.V. Bibikova and E.Y. Borisova. We thank Editor Mattia Pistone, reviewer Prof. Morishita and one anonymous reviewer for comments and suggestions, which helped to improve the manuscript. E. Sizova, J. Reimink, M. Belosevic, D. Baratoux and A.V. Sobolev are thanked for helpful discussion. We acknowledge the European Synchrotron Radiation Facility for providing access to beamtime, and O. Proux and M. Muñoz for help with XANES data acquisition and processing. C. McCammon and H. Keppler are thanked for offering access to the Bavarian Research Institute of Experimental Geochemistry and Geophysics (Germany). This work is funded by the AST "Planets" in Toulouse (France) and Deutsche Forschungsgemeinschaft (DFG, Germany) and ATUPS (Université Paul Sabatier, Toulouse, France) for providing a travel grant in 2015 – 2016 to N.R. Zagrtdenov and Russian Scientific Foundation (RSF) for partial support of the study (grant 18-17-00206) to O.G. Safonov.

## AUTHOR CONTRIBUTIONS

A.Y. Borisova, M.J.T. and O.G.S. developed the conceptual idea of the study; A.Y. Borisova and N.R.Z., S.S. prepared and conducted high T-P runs at BGI; N.R.Z., O.G.S., A.Y. Bychkov, A.T., V.M.P., D.A.V. prepared and performed experiments at IEM; N.R.Z., G.S.P., W.A.B, O.E.M. conducted thermodynamic modeling; G.S.P., N.R.Z. and A.A.G. carried out XAS and SIMS measurements, respectively; K.P.J., B.S. and U.W. performed LA-ICP-MS analyses. I.N.B. has performed oxygen isotope analyses of serpentinite. A.Y. Borisova, N.R.Z., S.G. P. de P. and K.E.A.C. performed microanalytical measurements and mapping using EPMA and A.Y. Borisova, A.N. and G.C. developed geological applications; all authors contributed to data interpretation and manuscript writing.

**FIGURES**

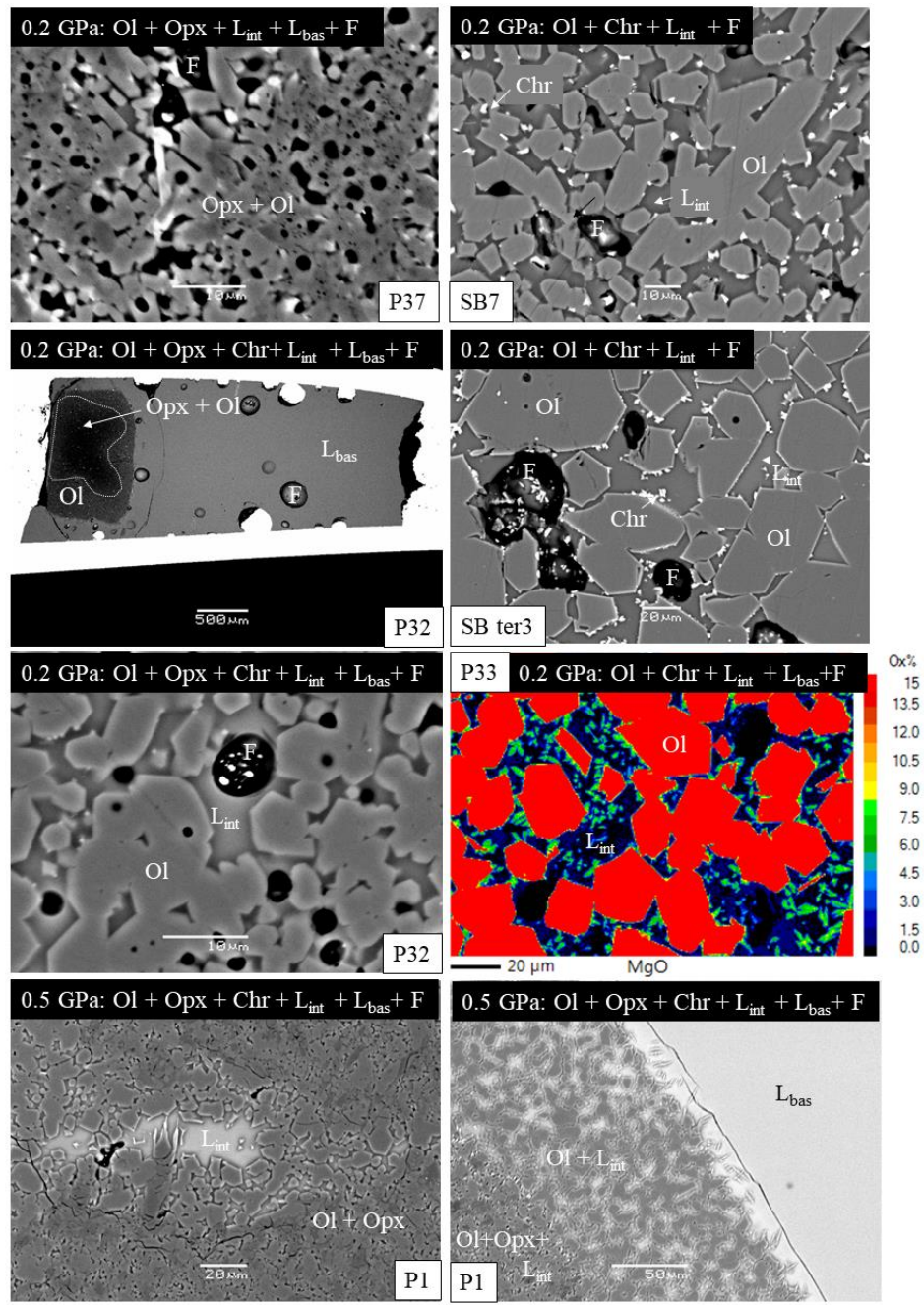

**Figure 1.** Back-scattered electron images and characteristic X-ray maps of the hybrid and mixed runs reflecting aqueous fluid-assisted partial melting of peridotite at 0.2 GPa and 0.5 GPa pressures. The images show mineral, glass and fluid phase associations observed in the capsules and, particularly, in the olivine-rich zones after the quench. Ol, Opx, Chr, $L_{int}$, $L_{bas}$, F on the images stand for olivine, orthopyroxene, chromite, interstitial and basaltic glasses, and fluid bubbles, respectively. The run numbers correspond to those used in **Supplementary Table A2**.

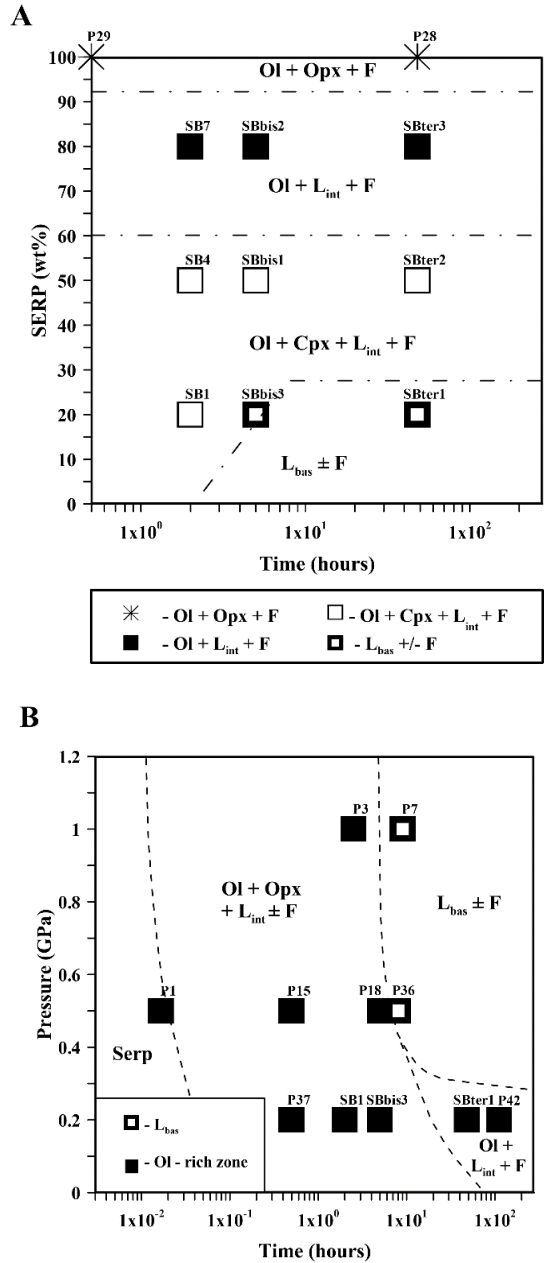

**Figure 2. (A)** Phase compositions plotted as serpentine (SERP) proportion (in wt%) versus run duration (in hours) for the pure serpentinite experiments and mixed serpentinite-basalt reaction experiments at 0.2 GPa and 1250°C. The diagram shows mineral, glass and fluid phase associations observed after quench. Ol, Opx, Cpx, $L_{int}$, $L_{bas}$, Chr, F stand for olivine, orthopyroxene, clinopyroxene, interstitial glass, basaltic glass and fluid, respectively. All olivine-rich samples demonstrate presence of chromite and chromiferous magnetite. The run numbers and the corresponding experimental products with their compositions are further detailed in **Supplementary Tables A2 – A4**. **(B)** Phase compositions plotted as pressure versus run duration (in hours) for the hybrid serpentinite-basalt reaction experiments at 0.2 to 1.0 GPa and 1250 – 1300°C. The diagram shows mineral, glass and fluid phase associations observed in the olivine-rich zone after quench. $L_{bas}$ is a unique zone of hydrous basaltic glass, and Ol-rich zone is olivine-rich zone with residual serpentine minerals (Serp), predominant magnesian olivine (Ol) and orthopyroxene (Opx) in interstitial glasses ($L_{int}$), frequently in the presence of fluid (F) bubbles. All olivine-rich samples demonstrate presence of chromite and chromiferous magnetite.

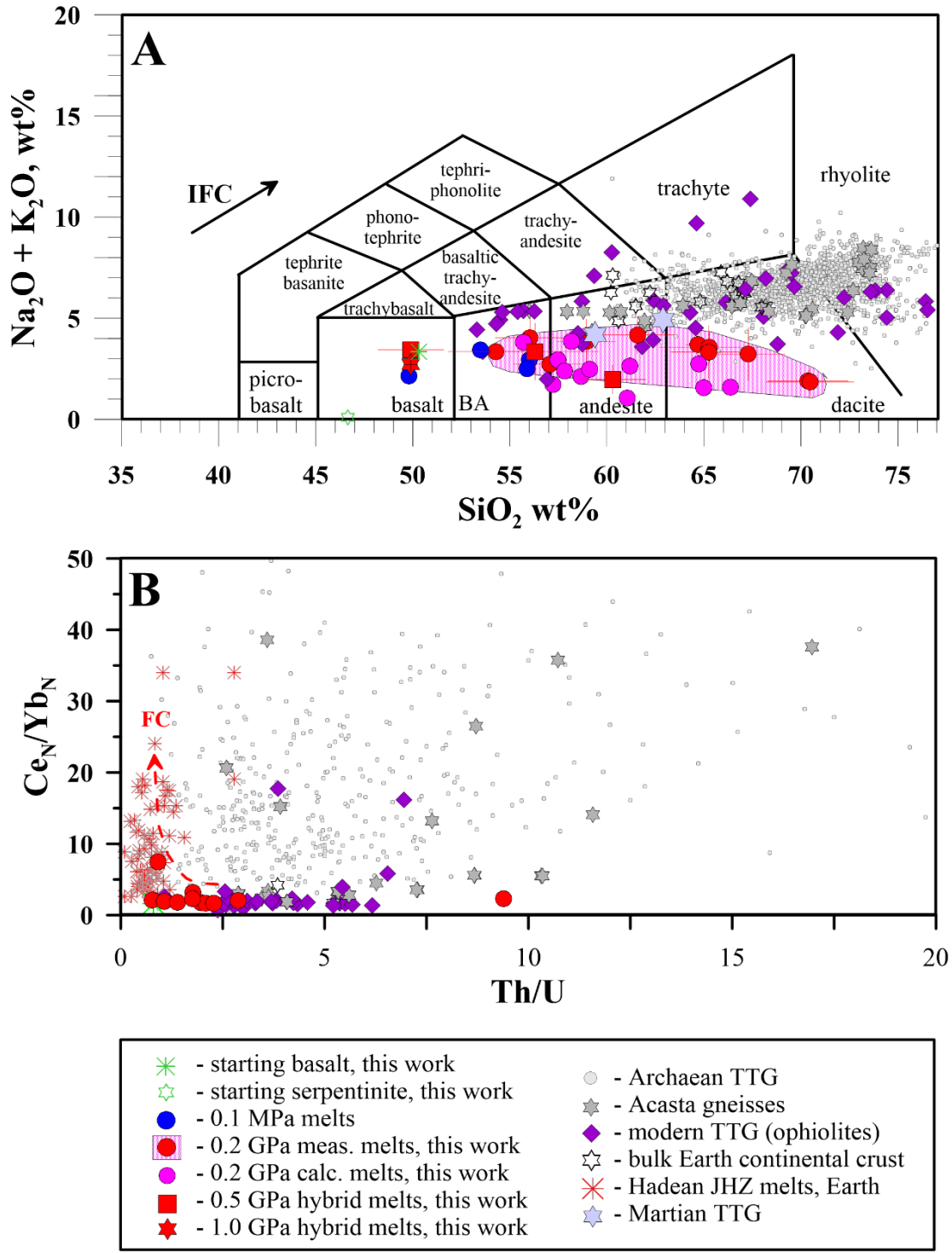

**Figure 3.** (A) Sum of alkali oxides (in wt%) versus $SiO_2$ (in wt%) in the interstitial glasses produced during hybrid and mixed runs of serpentinite-basalt interaction plotted in the classification diagram (Le Maitre, 2002). The major element composition of the measured ("meas. melts", red points contoured by purple ellipsoid) and recalculated ("calc. melts", marked as purple points) melts is recalculated to 100% on a volatile-free basis. The error bars reflect the felsic glass heterogeneity compared to the mean values. The glass composition is compared to the composition of the starting basalt and serpentinites, the anhydrous 0.1 MPa experimental glasses (Fisk, 1986), modern TTG (Amri et al., 1996; 2007; Shervais,

2008; Xu et al., 2017), the bulk Earth continental crust (Rudnick and Gao, 2003; Wedepohl, 1995), Archaean TTG (Moyen and Martin, 2012), Acasta Gneiss Complex rocks (Acasta gneisses) (Reimink et al., 2016), Martian TTG (Sautter et al., 2016), and melts in equilibrium with Jack Hills zircons at 800°C (Hadean JHZ melts, Earth) detailed in **Supplementary Table A9**. The IFC is isobaric fractional crystallization trend relating recalculated and measured melts. BA – basaltic andesite. **(B)** $Ce_N/Yb_N$ versus Th/U in the 0.2 GPa hybrid felsic glasses. For $Ce_N/Yb_N$, N denotes primitive mantle-normalized concentration after (Sun and McDonough, 1989). The red arrow demonstrates the tendency of $Ce_N/Yb_N$ and Th/U behavior upon the fractional crystallization (FC) of felsic magma.

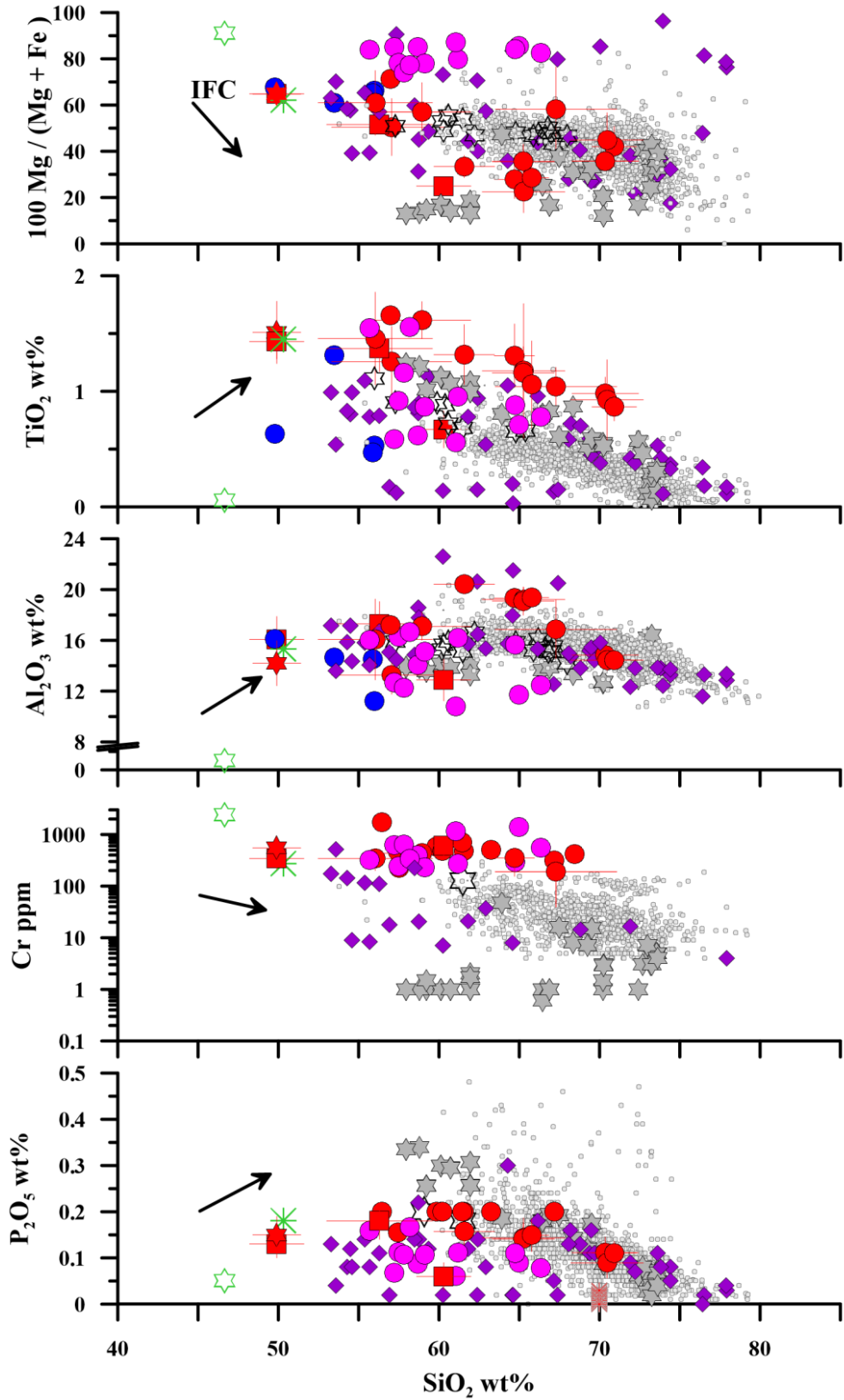
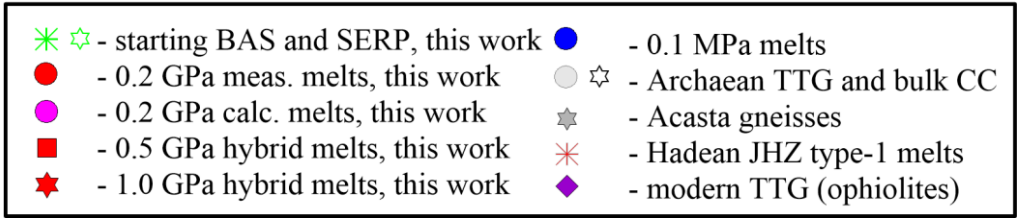

**Figure 4.** Magnesium index (100 Mg/(Mg+Fe)), major and minor Ti, Al, P oxides (in wt%), and Cr (in ppm) versus $SiO_2$ (in wt%) in the interstitial glasses produced at 0.2 GPa during serpentinite-basalt interaction of hybrid and mixed runs. The major element composition of the measured (meas.) and recalculated (calc.) melts is recalculated to 100% on a volatile-free basis. The error bars reflect the felsic glass inter-sample heterogeneity compared to the mean values (**Supplementary Tables A3, A4**). The glass composition is compared to the composition of the starting basalt (BAS) and serpentinite (SERP), the anhydrous 0.1 MPa experimental glasses (Fisk, 1986), modern TTG in mantle sections of ophiolites (Amri et al., 1996; 2007; Shervais, 2008; Xu et al., 2017), the bulk Earth continental crust (bulk CC) (Rudnick and Gao, 2003; Wedepohl, 1995), Archaean TTG (Moyen and Martin, 2012), Acasta Gneiss Complex rocks (Acasta gneisses) (Reimink et al., 2016), and melts in equilibrium with Jack Hills zircons at 800°C (**Supplementary Table A9**). The composition of the starting materials is given in **Supplementary Table A1**. The 70 wt% $SiO_2$ is suggested concentration in the melt responsible for the JHZ crystallization. Both EMPA and LA-ICP-MS data for the 0.2 GPa felsic melts are used for the Cr concentrations of hybrid melts, whereas only EPMA data are used for the P contents and LA-ICP-MS data for the trace elements (**Supplementary Table A3**). The IFC means isobaric fractional crystallization trend relating recalculated and measured melt compositions.

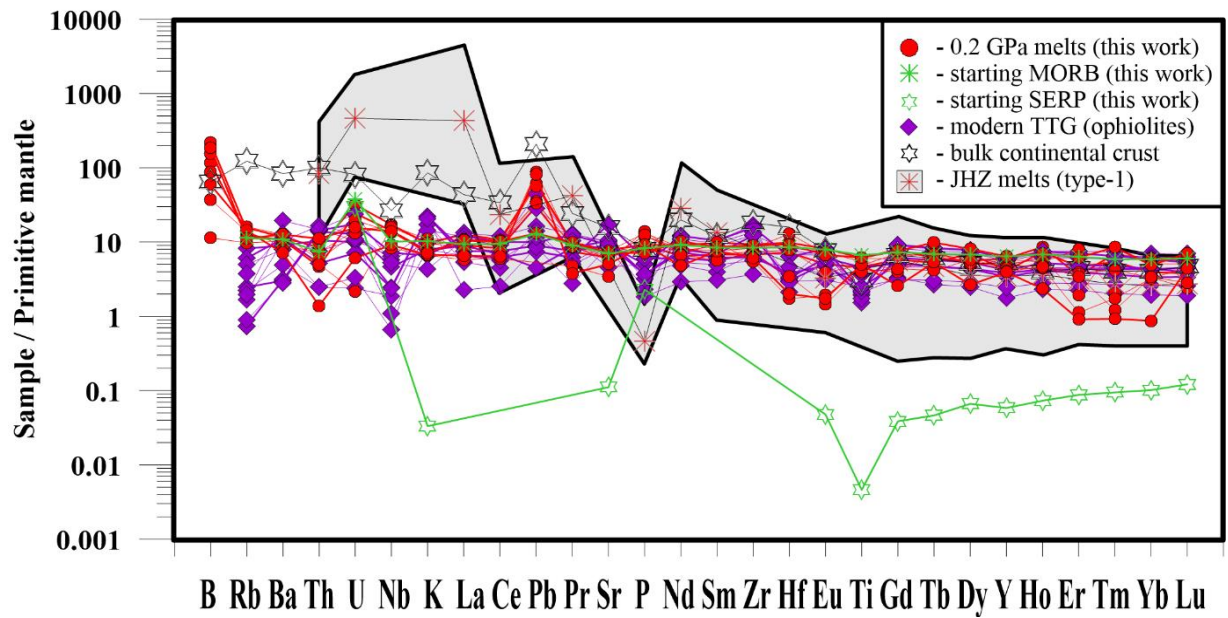

**Figure 5.** Primitive mantle-normalized patterns of trace element composition of felsic glasses produced at 0.2 GPa during hybrid runs of serpentinite-basalt interaction. The primitive mantle composition is after (Sun and McDonough, 1989) and (Lyubetskaya and Korenaga, 2007) for B only. The felsic glass composition is compared to that of the starting basalt (starting MORB) and serpentinite (starting SERP), modern TTG (Amri et al., 1996; 2007; Shervais, 2008; Xu et al., 2017), the bulk continental crust Wedepohl (1995), and calculated melts in equilibrium with Jack Hills zircons at 800°C (**Supplementary Table A9**). The grey field reflects the whole compositional range for melts in equilibrium with the Jack Hills zircons.

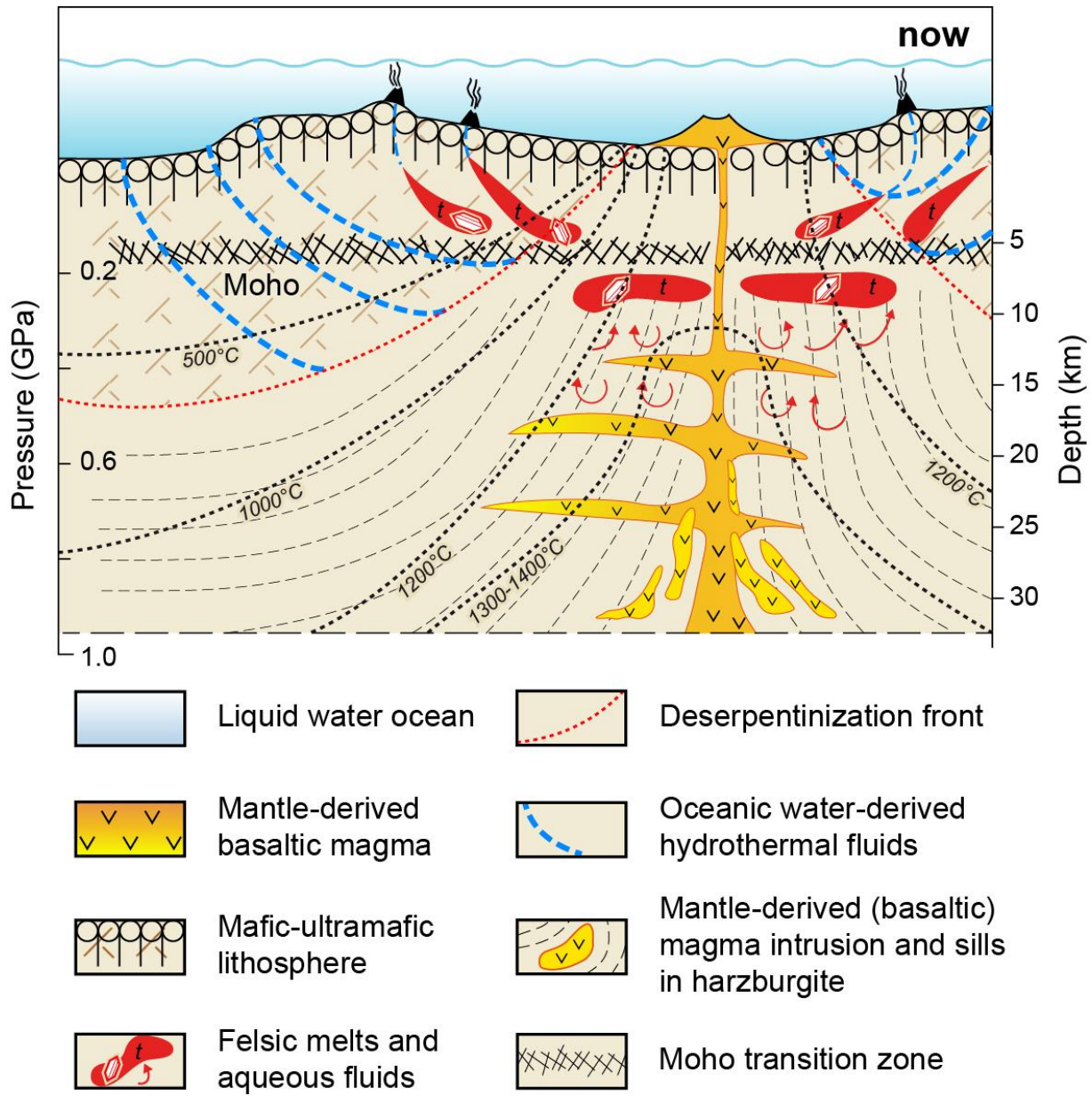

**Figure 6.** Model of formation of the modern shallow felsic crust in mantle peridotites of oceanic lithosphere. Intrusion of mantle-derived magmas into hydrated (serpentinized) mantle initiated the dehydration process. The released fluid promoted fertilization and partial melting of the peridotite. The partial melting produces TTG magmas marked as "t". The generated melts formed intrusion bodies in the host peridotite and overlaying crust at shallow <10 km depths. Black dotted lines correspond to isotherms. The horizontal scale of the model is ~30 km. The depths on the vertical scale are calculated based on a serpentinite density of 3 g/cm$^3$.

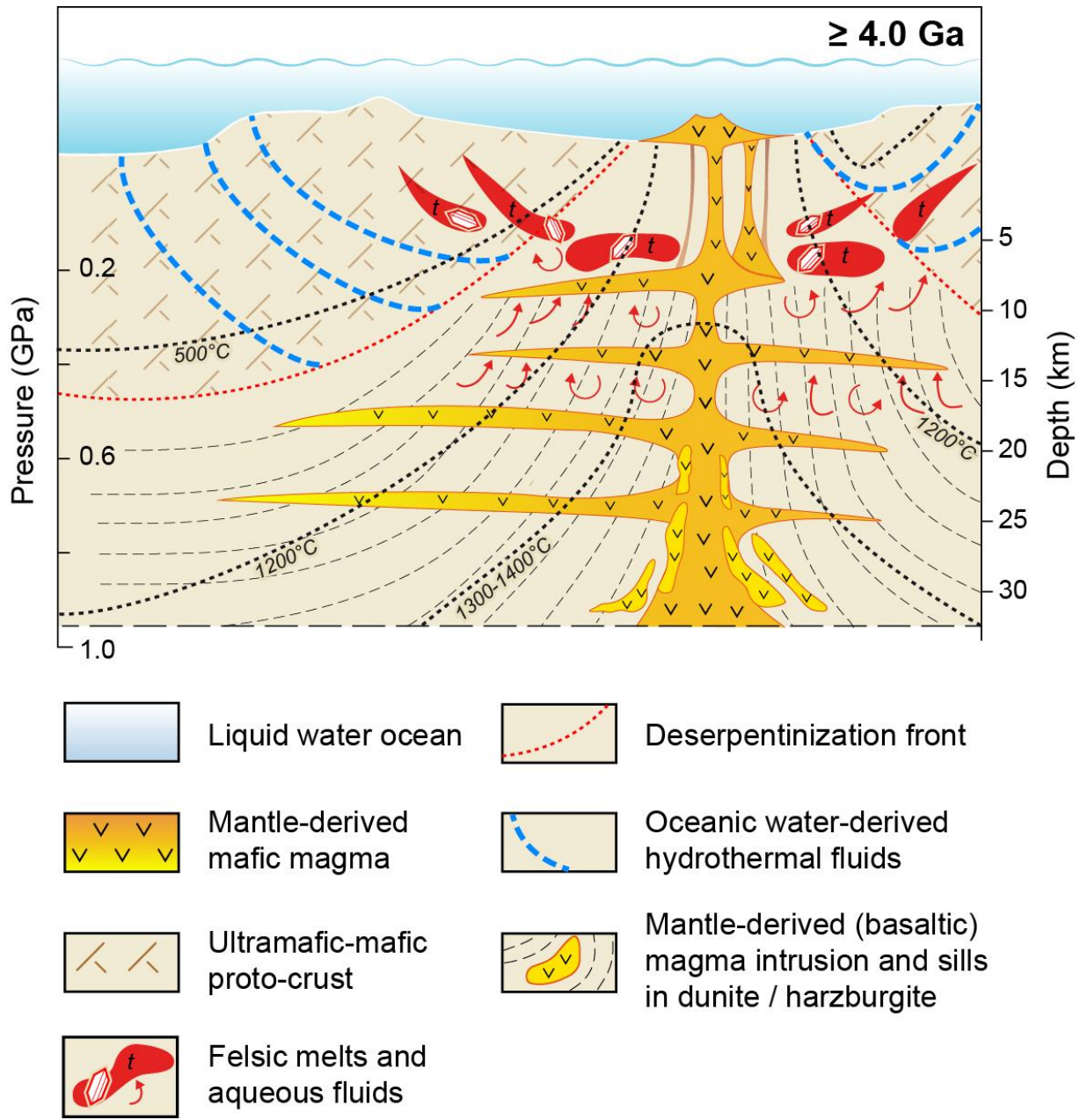

**Figure 7.** Model of formation of the Hadean shallow felsic crust produced due to activity of Hadean heat-pipe volcanoes. Intrusion of mantle-derived magmas into ultramafic-mafic protocrust initiated the dehydration process. The released fluid promoted fertilization and partial melting of the peridotite. The partial melting produces tonalite-granodiorite magmas marked as "t". The generated melts formed intrusion bodies in the host peridotite protocrust at shallow <10 km depths. Black dotted lines correspond to isotherms. The horizontal scale of the model is ~30 km. The depths on the vertical scale are calculated based on a serpentinite density of 3 g/cm$^3$.

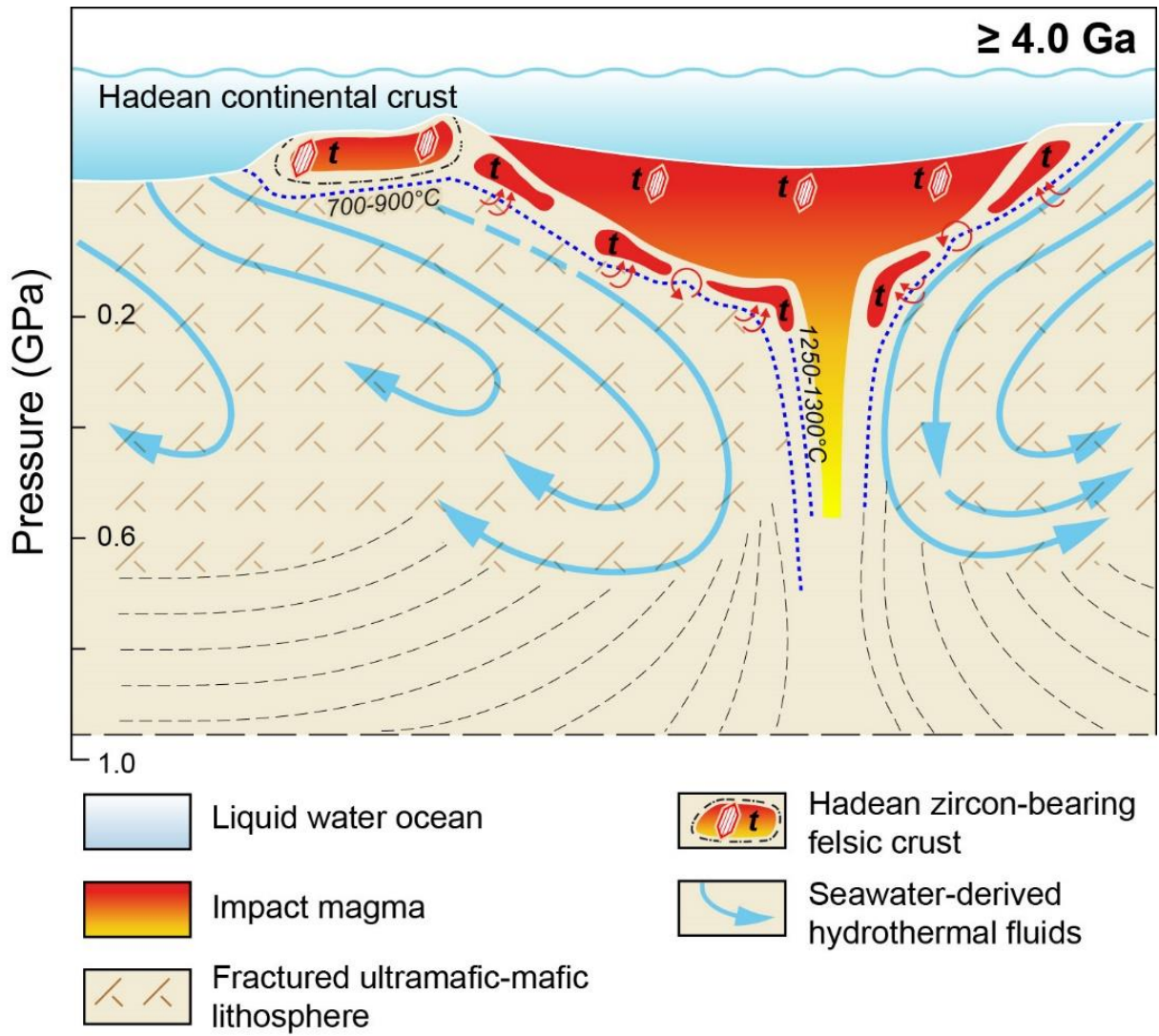

**Figure 8**. Model of impact formation of felsic crust on the early Earth and Mars. Impact-induced melting of the hydrated ultramafic-mafic protocrust promoted fertilization and partial melting of the peridotite. The partial melting produces intermediate to felsic (tonalite and granodiorite) magmas marked as "t". The intermediate to felsic melt production may happen at condition of the predominant proportion of serpentinite above 50 wt% into the hybrid system. The generated melts formed impact body at shallow <10 km depths. Blue dotted lines correspond to isotherms. The horizontal scale of the model is ~30 km.

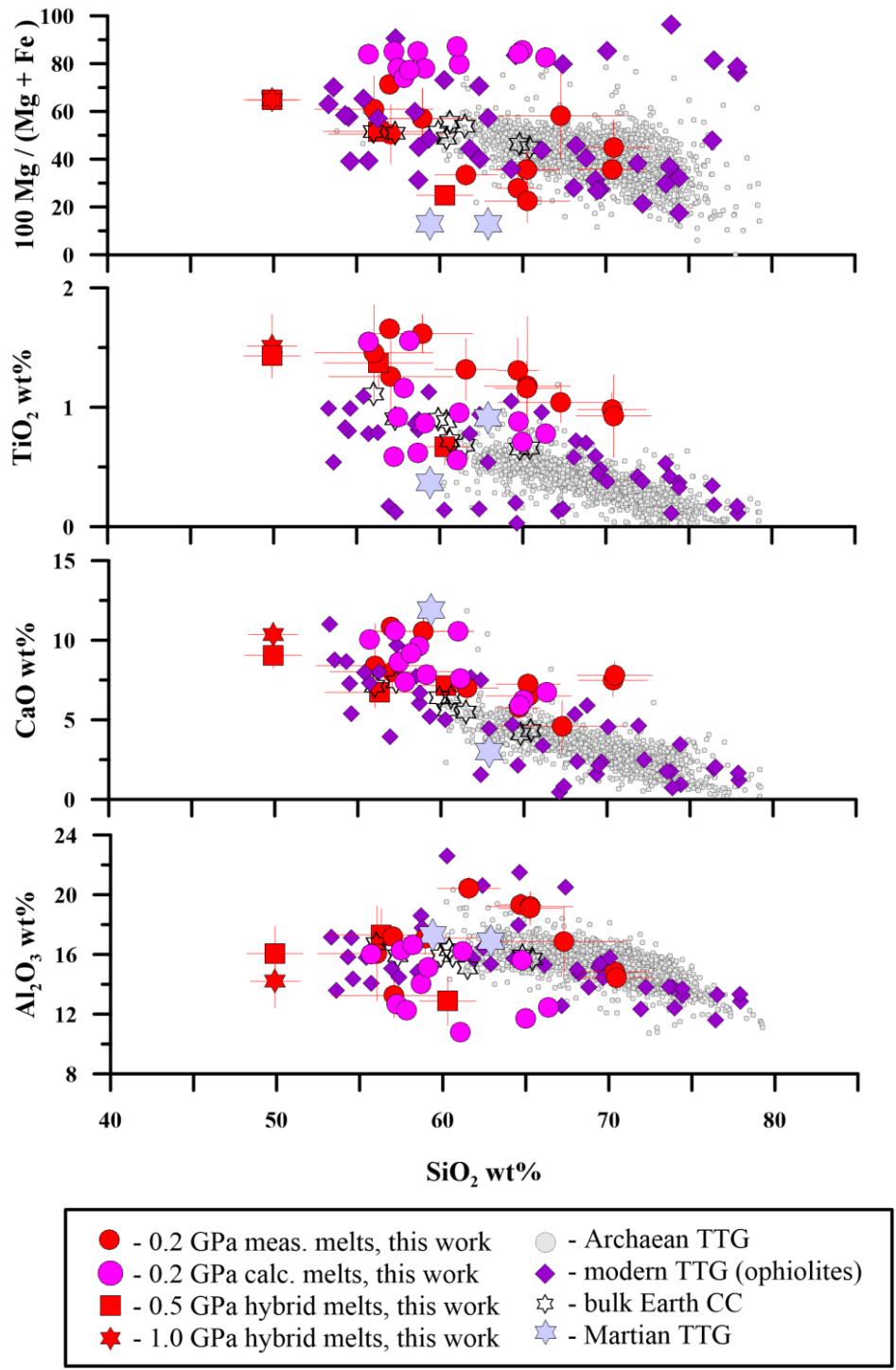

**Figure 9.** Magnesium index (100 Mg/(Mg+Fe)), major and minor $TiO_2$, CaO and $Al_2O_3$ oxides (in wt%), versus $SiO_2$ (in wt%) in the interstitial glasses produced at 0.2 GPa during serpentinite-basalt interaction of hybrid and mixed runs. The major element composition of the measured (meas.) and recalculated (calc.) melts is recalculated to 100% on a volatile-free basis. The error bars reflect the felsic glass heterogeneity compared to the mean values (**Supplementary Tables A3, A4**). The glass composition is compared to the modern mantle TTG (modern TTG) (Amri et al., 1996; 2007; Shervais, 2008; Xu et al., 2017), the bulk Earth continental crust (bulk Earth CC) (Rudnick and Gao, 2003), Martian TTG (Sautter et al., 2016), and Archaean TTG (Moyen and Martin, 2012).